\documentclass[fleqn,usenatbib]{mnras}

\usepackage{newtxtext,newtxmath,verbatim}

\usepackage{graphicx}
\usepackage{dcolumn}

\usepackage[usenames,dvipsnames]{xcolor}
\hypersetup{colorlinks=true,citecolor=blue,linkcolor=OliveGreen}

\usepackage{soul}

\usepackage{subfigure}
\usepackage{wasysym}
\usepackage{verbatim}
\usepackage{color}
\usepackage{mathtools}
\usepackage{hyperref}
\usepackage{soul}
\usepackage{pbox}
\usepackage[export]{adjustbox}

\usepackage{slashed}
\usepackage{multirow}
\usepackage{booktabs}  
\usepackage{natbib}
\usepackage{float}
\usepackage{afterpage}
\usepackage{dcolumn}
\usepackage{bm}
\usepackage{lipsum}
\usepackage{setspace}
\usepackage{etoolbox}

\def\beq{\begin{equation}}
\def\eeq{\end{equation}}

\def\mnras{Mon. Not. R. Astron. Soc}
\def\physrep{Physics Reports}

\usepackage[T1]{fontenc}
\usepackage{ae,aecompl}

\usepackage{graphicx}	
\usepackage{amsmath}	

\title[\textsc{ELGs in IllustrisTNG}]{The galaxy-halo connection of emission-line galaxies in IllustrisTNG}

\author[B. Hadzhiyska et al.]{
Boryana Hadzhiyska,$^{1}$\thanks{E-mail: boryana.hadzhiyska@cfa.harvard.edu}
Sandro Tacchella,$^{1}$
Sownak Bose,$^{1}$
and Daniel J. Eisenstein$^{1}$
\\
$^{1}$centre for Astrophysics $\vert$ Harvard \& Smithsonian, 60 Garden St., Cambridge, MA 02138, USA\\
}
\date{Accepted XXX. Received YYY; in original form ZZZ}

\pubyear{2019}

\begin{document}
\label{firstpage}
\pagerange{\pageref{firstpage}--\pageref{lastpage}}
\maketitle

\begin{abstract}
We employ the hydrodynamical simulation IllustrisTNG-300-1 to explore the halo occupation distribution (HOD) and environmental dependence of luminous star-forming emission-line galaxies (ELGs) at $z \sim 1$. Such galaxies are key targets for current and upcoming cosmological surveys. We select model galaxies through cuts in colour-colour space allowing for a direct comparison with the Extended Baryon Oscillation Spectroscopic Survey and the Dark Energy Spectroscopic Instrument (DESI) surveys and then compare them with galaxies selected based on specific star-formation rate (sSFR) and stellar mass. We demonstrate that the ELG populations are twice more likely to reside in lower-density regions (sheets) compared with the mass-selected populations and twice less likely to occupy the densest regions of the cosmic web (knots). We also show that the colour-selected and sSFR-selected ELGs exhibit very similar occupation and clustering statistics, finding that the agreement is best for lower redshifts. In contrast with the mass-selected sample, the occupation of haloes by a central ELG peaks at $\sim$20\%. We furthermore explore the dependence of the HOD and the auto-correlation on environment, noticing that at fixed halo mass, galaxies in high-density regions cluster about 10 times more strongly than low-density ones. This result suggests that we should model carefully the galaxy-halo relation and implement assembly bias effects into our models (estimated at $\sim$4\% of the clustering of the DESI colour-selected sample at $z = 0.8$). Finally, we apply a simple mock recipe to recover the clustering on large scales ($r \gtrsim 1 \ {\rm Mpc}/h$) to within 1\% by augmenting the HOD model with an environment dependence, demonstrating the power of adopting flexible population models.
\end{abstract}

\begin{keywords}
cosmology: large-scale structure of Universe -- galaxies: haloes -- methods: numerical -- cosmology: theory
\end{keywords}



\section{Introduction}
\label{sec:intro}
In the current cosmological paradigm, the Universe is made up of a 
dense network of filaments shaped by gravity. 
Embedded in these filaments are dark matter structures, called
haloes, which correspond to overdense regions that have evolved by gravitational
instability and interactions with other haloes. According to this framework,
galaxy formation takes place within these haloes. Baryonic matter sinks to the
centre of their gravitational potential wells and condensation of cold gas 
allows for galaxies to form and evolve
\citep{1978MNRAS.183..341W}. A detailed understanding of the galaxy-halo connection 
would enable us to use the galaxy field to place stringent constraints
on cosmological parameters. 

The evolution and distribution of the dark matter
haloes can be modeled effectively using cosmological ($N$-body) simulations, 
which incorporate various assumptions of the cosmological model. These simulations
have the benefit of encompassing very large volumes ($\sim$1 Gpc$/h$),
but they lack prescriptions for determining the evolution and distribution 
of the baryonic content.
To probe the effect of baryons in such large volumes, cosmologists often adopt
alternative schemes for populating the dark matter haloes with galaxies through 
\textit{a posteriori} models of varying complexity. The simplest of them
are phenomenological approaches such as halo occupation distribution (HOD) and
subhalo abundance matching (SHAM), which rely on basic assumptions of how galaxies are connected to their haloes. 

The HOD framework \citep{2000MNRAS.311..793B,2000MNRAS.318.1144P,2001ApJ...546...20S,
2004MNRAS.350.1153Y} provides an
empirical relation between halo mass and the number
of galaxies it hosts, which is expressed as the probability 
distribution $P(N|M_h)$ that a halo of virial mass $M_h$ hosts
$N$ galaxies satisfying some selection criteria. This method can thus
be used to study galaxy clustering \citep{HOD,Zheng:2004id,
2006ApJ...647..201C,2011ApJ...736...59Z,2018ARA&A..56..435W}. Since the HOD
parameters are tuned so as to
reproduce only a limited set of observables such as the two-point correlation function and the galaxy number density,
HOD modelling is one of the most efficient ways to 
``paint'' galaxies on top of $N$-body simulations of 
large volumes and produce the many realizations required for, e.g., 
estimating covariance matrices using mock galaxy catalogues 
\citep[e.g.][]{2009MNRAS.396...19N,2013MNRAS.428.1036M}. Such mock catalogues
are crucial for developing new algorithms that will be used 
for the next generation of surveys such as DESI and \textit{Euclid}.
However, in their most stripped-down versions, empirical approaches
such as the mass-only HOD have been shown to lead to significant
discrepancies with observations and more detailed galaxy
formation models, so one should proceed with caution when adopting
them \citep{2007MNRAS.374.1303C,2015MNRAS.454.3030P,2019arXiv190811448B}. For example, several recent works have shown
that local halo environment as a secondary HOD parameter yields 
a better agreement with simulations than mass-only
prescriptions \citep{2020MNRAS.493.5506H,2020MNRAS.492.2739X,2020arXiv200804913H}.

Alternatively, other more complex empirical models that study the relationship between the properties of galaxies and their host haloes have been an area of keen interest \citep{2013ApJ...770...57B,2013ApJ...768L..37T, 2013MNRAS.428.3121M,2018MNRAS.477.1822M,2018ApJ...868...92T,2019MNRAS.488.3143B}. A particularly well-developed route for assigning galaxies to dark-matter haloes is through semi-analytical models of the galaxy formation (SAMs). SAMs provide
a physically motivated mechanism for evolving galaxies and studying the processes
that shape the evolution of baryons. These are built on top of haloes extracted from 
an $N$-body simulation and employ various physical prescriptions to make predictions
for the abundance and clustering of galaxies \citep{2000MNRAS.319..168C,
2006RPPh...69.3101B,2015ARA&A..53...51S}. A disadvantage
to these models is that their outputs are highly sensitive to the 
particular choices of calibration parameters, which in turn depend on many
physical processes that are still poorly understood.

Hydrodynamical simulations, on the other hand, are an example
of an \textit{ab initio} approach to gaining insight into the formation and 
evolution of galaxies \citep[e.g.][]{2014MNRAS.444.1518V,2015MNRAS.446..521S},
incorporating baryonic effects such as stellar wind,
supernova feedback, gas cooling, and black hole feedback. While these
models are computationally expensive and cannot (at present) be run over the large volumes needed
for future cosmological galaxy surveys, they can still be used to inform us of the galaxy-halo
connection on large scales, having recently reached sizes of several hundred
megaparsecs \citep{2016MNRAS.460.3100C,2018MNRAS.475..676S}.
In particular, we can test the
accuracy of empirical approaches such as the HOD model by applying them to the dark-matter-only
counterpart of a hydro simulation and comparing various statistics
of the galaxy field \citep{sownak,2020MNRAS.493.5506H,2020arXiv200503672C}. Such analyses
have been performed extensively for galaxy populations 
above a certain stellar mass cut, 
but not as much for the study of other galaxy populations 
targeted by future surveys such as emission-line galaxies (ELGs).
Both observations and models indicate that star-forming
galaxies in general, and ELGs in particular, populate 
haloes in a different way than mass-selected samples
\citep[e.g.][]{Zheng:2004id,2018MNRAS.480..864C,2017MNRAS.472..550F,
2016MNRAS.459.3040G,2018MNRAS.474.4024G,2019MNRAS.483.4501A}.

Emission-line galaxies (ELGs), targeted for their sensitivity
to the Universe expansion rate during the epoch dark energy
starts to dominate the energy budget of the Universe ($z \sim 1$), are characterized by nebular
emission lines produced by ionized gas in the interstellar medium. 
The gas gets heated either by newly formed stars, evolved stars, or by
nuclear activity due to black hole accretion.
In this work, we focus only on luminous star-forming ELGs with a high line luminosity, as this is
the population targeted by many current cosmological surveys
such as DESI, \textit{Euclid}, SDSS-IV/eBOSS, and the \textit{Nancy Grace Roman Space Telescope}
\citep{2013arXiv1308.0847L,2018LRR....21....2A,2018MNRAS.473.4773A,2015arXiv150303757S}. 
The presence and strength of the emission lines depends
on a number of factors, including star
formation rate (SFR), gas metallicity, and particular 
conditions of the HII regions \citep[e.g.][]{2014MNRAS.443..799O}. 
Despite the fact that ELG samples are related to star 
formation, they are not equivalent to SFR-selected samples. 
Nevertheless, the HOD method can still be adopted to study them 
\citep{2012MNRAS.426..679G,2017MNRAS.469.2913C,2018MNRAS.480..864C,
2020arXiv200709012A}. For both galaxy populations,
the shape of the HOD is more complex than that
of the more widely studied stellar-mass
selected samples \citep{2013MNRAS.432.2717C,2018MNRAS.474.4024G}. 
For example, the occupation function of 
central galaxies does not follow a step-like form but rather
peaks at low halo-mass values and decays promptly after.

Understanding the connection between ELGs and their host dark 
matter haloes will help us obtain more realistic mock catalogues,
which are crucial for the analysis of future observational samples.
Observations concentrating on star-forming ELGs around $z \sim 1$
have studied their occupancy as a function of halo mass
\citep{2016MNRAS.461.3421F,2018MNRAS.478.2999K,2019ApJ...871..147G}
as well as their clustering bias \citep{2013MNRAS.428.1498C,2020arXiv201008500J}.
There have been efforts to match these findings through 
SAMs \citep{2018MNRAS.474.4024G,2017MNRAS.472..550F,2020MNRAS.498.1852G}, which 
despite being able to recover the occupation function, 
have found the clustering to be inconsistent with observations on 
large scales. A possible explanation is the lack of ``assembly bias'' in 
the models interpreting the observations \citep{2019MNRAS.484.1133C}, 
where we define ``assembly bias'' to be the additional properties
of haloes which affect the galaxy clustering beyond halo mass. 
Another set of assumptions that these results rely upon is the
particular calibration scheme adopted by the SAM. It is, therefore,
worth exploring these questions in hydrodynamical simulations,
which implement detailed baryonic physics processes during the
simulation run instead of once the run has completed assuming
gravitation-only interactions.

In this work, we study the clustering, galaxy bias and occupation function of
ELG samples obtained by applying the colour-space selection criteria
proposed by DESI and eBOSS to the output of the hydrodynamical 
simulation IllustrisTNG \citep{2018MNRAS.473.4077P,2019ComAC...6....2N}. We further construct a mock catalogue
using the HOD model augmented with an environmental dependence
in order to recover the two-point correlation function on
large scales. In Section \ref{sec:meth}, we introduce
the DESI and eBOSS galaxy surveys, the hydrodynamical
simulation IllustrisTNG, and a description of the
steps we take to synthesize the galaxy colours. In Section \ref{sec:res},
we show the main results of our analysis. We first compare the colour-selected
ELGs with galaxy samples selected based on (s)SFR in terms of
their three-dimensional distributions in ($g-r$)-($r-z$)-(s)SFR space. We
then study their spatial distribution in the cosmic web, comparing 
it with that of a stellar mass-selected sample, with which we aim to mimic a luminous red galaxy (LRG) sample. We confirm that ELGs are more likely to
reside in filamentary structures. We then derive and study in detail
the occupation distribution
and auto-correlation function of both the sSFR- and colour-selected ELG-like samples and 
investigate their differences and similarities. In particular, we explore the dependence of 
both statistics on the density of the local environment. Finally in Section \ref{sec:mock}, we propose an
HOD prescription augmented with a secondary environment parameter, which captures well the clustering across all scales.
In Section \ref{sec:disc}, we summarize our results and comment on
their implications for future galaxy surveys.
\section{Methodology}
\label{sec:meth}
\begin{table*}
\caption{colour-space cuts applied to the model galaxies in order to mimic the selection in the corresponding observational survey. The eBOSS selection is taken from \citet{2017MNRAS.471.3955R}, while the DESI selection is as stated in the Final Design Report \citep{2016arXiv161100036D}. All magnitudes are reported in the AB system. The filter response used for the different cuts corresponds to the DECam camera.}
\begin{center}
\begin{tabular}{|c|c|c|c|}
\hline
Galaxy survey & Magnitude limit & Magnitude cut & colour selection\\
\hline \hline \vspace{0.0cm}

eBOSS-SGC  & $g < 24.7 \, \& \, r < 23.9 \, \& \, z < 23.0$ &  $21.825< g < 22.825$ & $-0.068(r-z) + 0.457 < (g-r) < 0.112(r-z) + 0.773$  \& \\ 
     & & & $0.218(g-r) + 0.571 < (r-z) < -0.555(g-r) + 1.901$ \\  
\hline
DESI & $g < 24.0 \, \& \, r < 23.4 \, \& \, z < 22.5$ & $20.0 < g < 23.6$  & $0.3 < (r-z) < 1.6$  \&  \\
     & & & $(g-r) < 1.15\cdot(r-z) - 0.15$ \& $(g-r) < -1.2\cdot(r-z) + 1.6$ \\
\hline
\end{tabular}
\end{center}
\label{tab:obs}
\end{table*}

In this section, we describe the two ELG target selection
strategies employed by the DESI and eBOSS surveys,
the hydrodynamical simulation IllustrisTNG (TNG) used
in this study, as well as the procedure we follow to 
synthesize the colours of the TNG galaxies. We also
present an alternative ELG selection approach
based on sSFR and a single photometric band
cut, which mimics the logic of the cuts in colour-colour space.

\subsection{DESI}
\label{sec:meth.desi}
DESI is a ground-based experiment that aims to place stringent constraints on
our models of dark energy, modified gravity and inflation, as well as the neutrino
mass by studying baryon acoustic oscillations (BAO) and the growth of structure 
through redshift-space distortions (RSD). The DESI instrument will conduct a five-year
survey of the sky designed to cover 14,000 deg$^2$.
To trace the underlying dark matter distribution, several classes of spectroscopic samples
will be selected including luminous red galaxies (LRGs), emission-line galaxies
(ELGs), quasars, and bright galaxies around $z \approx 0.2$, totaling more than
30 million objects. The ELG sample will probe the Universe out to redshifts of $z = 1.6$, targeting galaxies with bright [O II] emission lines 
\citep{2016arXiv161100036D}. 

The ELG DESI targets are selected using optical $grz$-band photometry from the Dark Energy Camera Legacy Survey (DECaLS), the Beijing-Arizona Sky Survey (BASS), and the Mayall z-band Legacy Survey (MzLS) \citep{2017PASP..129f4101Z,2019AJ....157..168D,2020arXiv200205828B}. 
The imaging depth is at least 24.0, 23.4, 22.5 AB (5$\sigma$ for an exponential 
profile $r_{\rm half-light}= 0.45"$) in $g$, $r$, and $z$.
The Final Design Report \citep{2016arXiv161100036D} colour cuts are listed in Table \ref{tab:obs}.

\subsection{eBOSS}
\label{sec:meth.eboss}
eBOSS is one of three surveys from the SDSS-IV experiment
\citep{2017AJ....154...28B}. It focuses on four different tracers,
which significantly expand the volume covered by BOSS,
spanning a redshift range of $0.6 < z < 2.2$. The four tracers are
LRGs \citep{2016ApJS..224...34P}, ELGs \citep{2017MNRAS.471.3955R}, 
`CORE' quasars \citep{2015ApJS..221...27M}, 
and variability-selected quasars \citep{2016A&A...587A..41P}. 
The 255 000 ELG targets are centered around $z \sim 0.8$ and are observed
through 300 plates with the BOSS spectrograph, obtaining
a $\sim$2\% precision distance estimate. 
The eBOSS ELG footprint is divided into two regions: South Galactic Cap
(SGC) and North Galactic Cap (NGC), covering 
$\sim$620 ${\rm deg^2}$ and $\sim$600 ${\rm deg^2}$.

The eBOSS ELG target selection is based on an earlier
data release of the ongoing imaging survey DECaLS \citep{2017MNRAS.471.3955R,2017MNRAS.465.1831D}. The cuts applied in colour-colour space to obtain the ELG sample
are shown in Table \ref{tab:obs}.

\subsection{IllustrisTNG}
\label{sec:meth.tng}
For modeling the galaxy population, 
we consider the hydrodynamical simulation IllustrisTNG \citep[TNG,][]{2018MNRAS.475..648P,2018MNRAS.480.5113M,2018MNRAS.477.1206N,
2018MNRAS.475..676S,2019ComAC...6....2N,2018MNRAS.475..624N,2019MNRAS.490.3196P,2019MNRAS.490.3234N}. 
TNG is a suite of cosmological magneto-hydrodynamic simulations, 
which were carried out using the \textsc{AREPO}
code \citep{2010MNRAS.401..791S} with cosmological parameters consistent with the
\textit{Planck 2015} analysis \citep{2016A&A...594A..13P}.
These simulations feature a series of improvements
compared with their predecessor, Illustris, such as
improved kinetic AGN feedback and galactic wind
models, as well as the inclusion of magnetic fields.

In particular, we will use IllustrisTNG-300-1
(TNG300 thereafter), the largest high-resolution hydrodynamical simulation from the suite. 
The size of its periodic box is 205 Mpc$/h$ with 2500$^3$ DM particles
and 2500$^3$ gas cells, implying a DM particle mass of $3.98 \times 10^7 \ \rm{M_\odot}/h$ and
baryonic mass of $7.44 \times 10^6 \ \rm{M_\odot}/h$. 
We also employ its DM-only counterpart, TNG300-Dark, which was evolved with the 
same initial conditions and the same number of dark matter particles ($2500^3$), each with 
particle mass of $4.73 \times 10^7 \ \rm{M_\odot}/h$. The haloes (groups) in TNG are found with a standard 
friends-of-friends (FoF) algorithm with linking length $b=0.2$ (in units of the mean interparticle spacing)
run on the dark matter particles, while the subhaloes are identified 
using the SUBFIND algorithm \citep{Springel:2000qu}, which detects 
substructure within the groups and defines locally overdense, self-bound particle groups.
We analyse the simulations at redshifts $z = 0.8$, $z = 1.1$ and $z = 1.4$,
considering galaxies with minimum stellar mass of $M_{\star} = 10^{10} M_\odot$, 
which are deemed to be well-resolved (i.e. more than 1000 stellar particles).

\subsection{Modeling galaxy colours}
\label{sec:meth.fsps}
\subsubsection{Stellar population synthesis and dust model}

For all galaxies in a given snapshot, we predict the DESI and eBOSS photometries ($g$-, $r$- and $z$-band) using the Flexible Stellar Population Synthesis code \citep[FSPS,][]{2010ascl.soft10043C,2010ApJ...712..833C}. We adopt the MILES stellar library \citep{2015MNRAS.449.1177V} and the MIST isochrones \citep{2016ApJ...823..102C}. We measure the star-formation history (SFH) in the simulation from all the stellar particles in a subhalo within 30 kpc of its center. We split the star-formation history (SFH) of each galaxy into a young (stellar ages $<30$ Myr) and old (stellar ages $>30$ Myr) component. We use the young SFH component to predict the nebular continuum emission and emission lines, assuming the measured gas-phase metallicity from the simulation and --1.4 for the ionization parameter (fiducial value in FSPS). We feed the old SFH component along with the mass-weighted stellar metallicity history to FSPS in order to predict the stellar continuum emission. For the galaxies studied here, the latter component dominates the flux in the DESI photometric bands. 

An additional key ingredient in predicting the photometry is the attenuation by dust. There are several different ways of how to model dust attenuation in the simulation \citep[e.g.][]{2018MNRAS.475..624N,2020MNRAS.492.5167V}. We use an empirical approach by basing our attenuation prescription on recent observational measurements. Specifically, we assume that the absorption optical depth follows:

\begin{equation}
    \tau_{\rm v} = \gamma \left(\frac{Z_{\rm gas}}{Z_{\odot}}\right)^{\alpha} \tilde{\Sigma}_{\star}^{\beta},
\end{equation}

\noindent
where $Z_{\rm gas}$ is the gas-phase metallicity and $\tilde{\Sigma}$ is the normalized stellar mass density ($\tilde{\Sigma}=\Sigma_{\star}/\langle\Sigma_{\star}\rangle$ with $\Sigma_{\star}=M_{\star}/(\pi r_{\rm e}^2$), where $r_{\rm e}$ is the half-mass radius). Both quantities are  obtained directly from the simulations. The parameters $\alpha$, $\beta$ and $\gamma$ have been roughly tuned to reproduce observed $z\sim0$ scaling relations between $\tau_{\rm v}$, SFR and $M_{\star}$ by \cite{2018ApJ...859...11S}, which is based on GALEX, SDSS and WISE photometry. Specifically, we find $\alpha$, $\beta$ and $\gamma$ to be $-0.6$, $0.2$, $0.4$, respectively. We also vary the additional dust attenuation toward younger stars (\texttt{dust1} in FSPS) and the dust index (shape of the dust attenuation law), and find that values close to the standard values within FSPS reproduces well the observed colour distribution at the redshifts of interest (shown in Section \ref{subsubsec:colour_comparison}). We emphasize that the dust model parameters are only roughly tuned to observations and we did not formally optimize the dust model parameters.

\subsubsection{Adding photometric uncertainties}
To make the synthetic sample more observationally
realistic, we add photometric scatter to each of the $grz$ bands. At the faint magnitudes we are working with, 
the noise is limited by the background rather than
the source, so the scatter can be assumed constant, i.e. independent of the source.  

To determine the amount of scatter we need to add, we first convert the 5$\sigma$ limiting magnitudes to fluxes
and divide by 5 to get the RMS. This step is often
complicated by the fact that the 5$\sigma$ limit depends on
the angular size of the galaxy -- point sources will have a 
fainter limit than extended sources.  For that purpose,
the Legacy Survey quotes the depths in terms of the 
standard reference galaxy (0.45" half-light exponential). We then
set the ratio between the three bands to
be constant and empirically scale the scatter until we
match the width of the colour distribution in the data.

\begin{figure}
\centering  
\includegraphics[width=.42\textwidth]{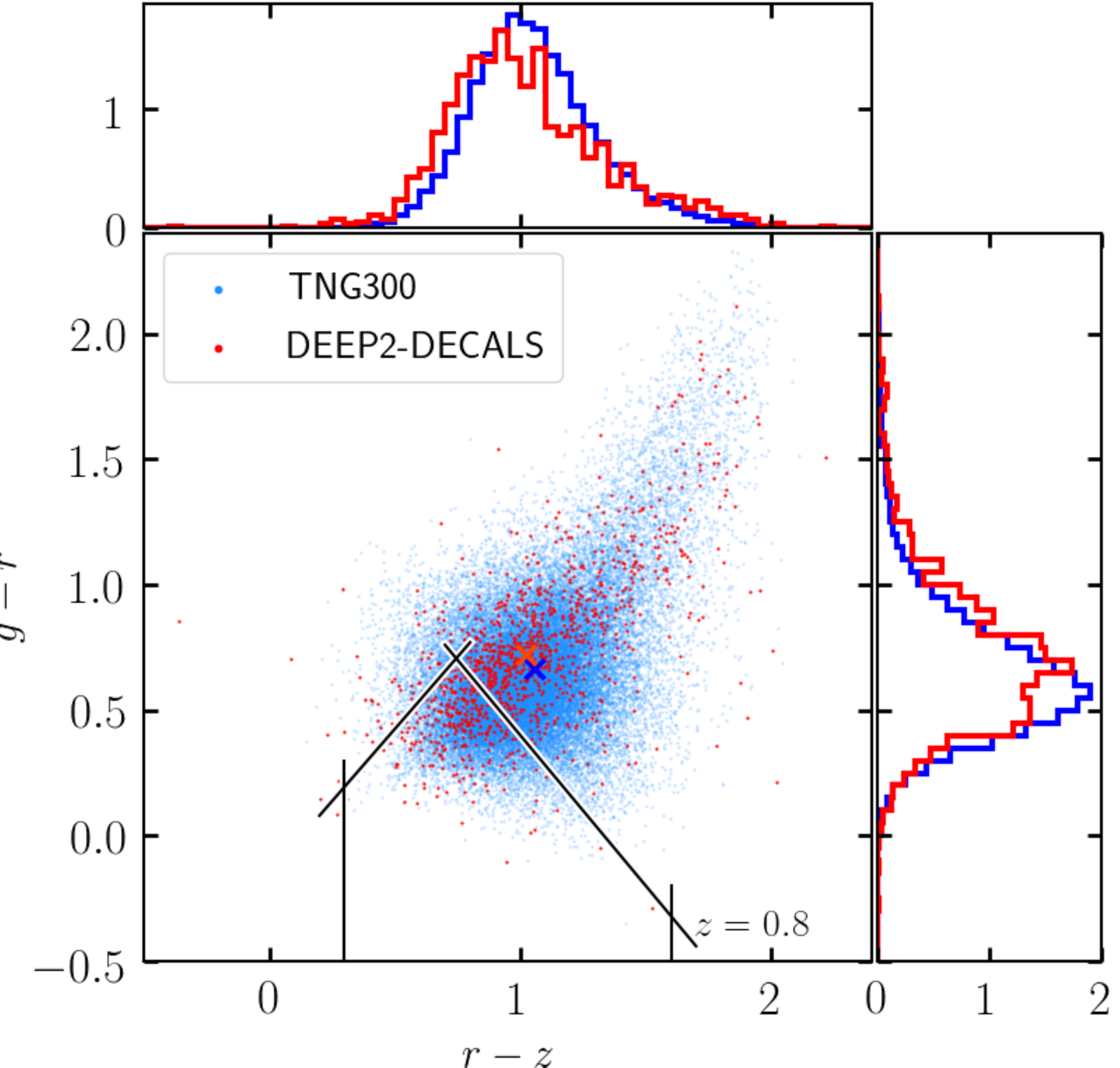}
\includegraphics[width=.42\textwidth]{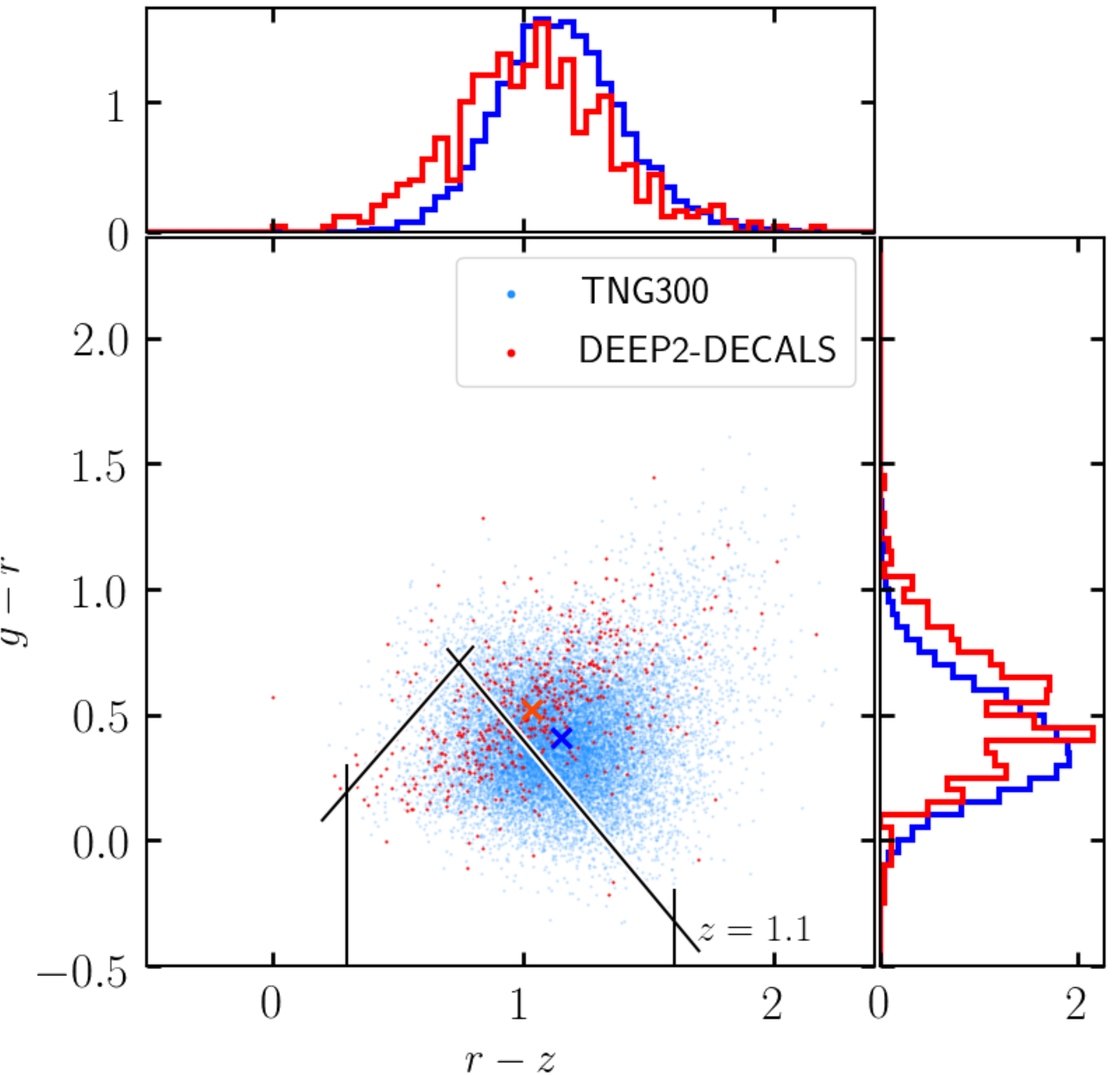}
\includegraphics[width=.42\textwidth]{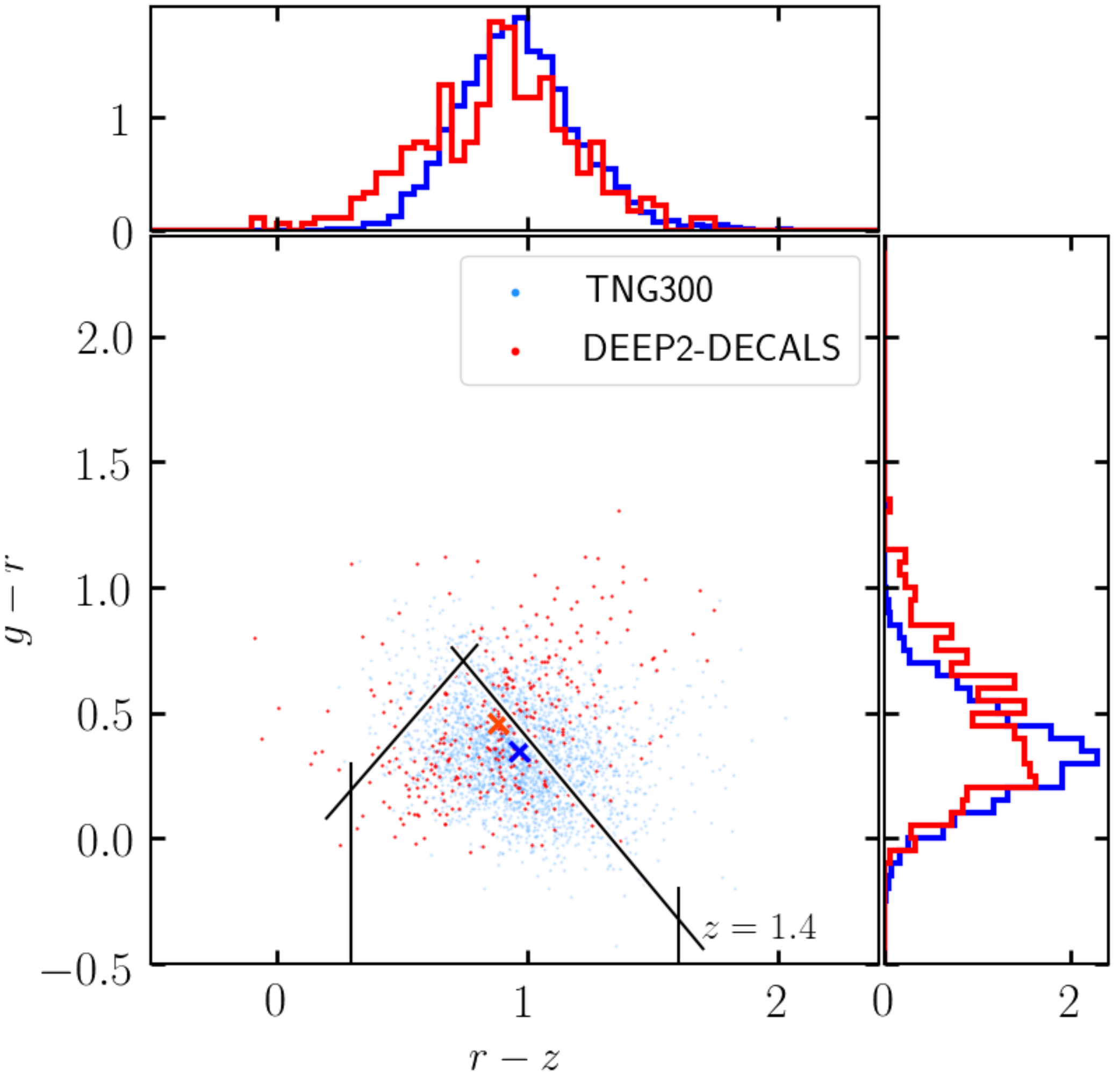}
\caption{Scatter plot of the colour-colour distribution of the synthesized TNG galaxy population (\textit{in blue}) and the DEEP2-DECaLS DR8 cross-matched catalogue (\textit{in red}). The three panels correspond to $z = 0.8$, $z = 1.1$, and $z = 1.4$ and show the $g-r$ vs. $r-z$ distributions. We apply DESI magnitude limit cuts from Table~\ref{tab:obs} to both samples. The location of the two loci (and the width of the distributions) is very similar at $z = 0.8$ sample, and differs by about $0.1$ mag for the other two redshift samples. The TNG-predicted SEDs are fainter in the $r$-band for a given $g-z$ colour than the observations, yielding bluer $g-r$ and redder $r-z$ colours.}
\label{fig:deep2}
\end{figure}

As will be discussed next, it turns out that the observed
scatter can be matched in a satisfactory manner even without
having to perform the second step of empirically matching
the scatter (see Fig.~\ref{fig:deep2}).

\subsubsection{Comparison of colour distribution with DEEP2-DECaLS}
\label{subsubsec:colour_comparison}
The DEEP2 Galaxy Redshift Survey \citep{2013ApJS..208....5N}
is a large spectroscopic redshift survey. The survey was conducted on the Keck II telescope using
the DEIMOS spectrograph \citep{2003SPIE.4841.1657F}
and measured redshifts for $\sim$32000 galaxies from $0<z<1.4$.
The DEEP2 targets were selected via a combination of magnitude  
($R_{\rm AB} <24.1$) and colour cuts ($BRI$) to efficiently select
galaxies with $z>0.7$. The DEEP2 survey has been particularly 
relevant for DESI, as its data has been used for
calibration and refinement of the DESI target selection.

Here, we perform cross-matching between the DECaLS data
release 8 (DR8) and DEEP2 in order to compare
the synthesized TNG colours at $z = 0.8$, $z = 1.1$ and $z = 1.4$ 
with the observed colours and examine the dust parameters
in the FSPS model. The matching is performed by first applying
the DEEP2 window functions and the 
Tycho-2 mask\footnote{The Tycho-2 catalogue is an astronomical reference catalogue of more than 2.5 million of the brightest stars. For more information, see \url{www.cosmos.esa.int/web/hipparcos/tycho-2}.} and then bijectively matching 
the DEEP2 and DR8 catalogues while testing for any astrometric discrepancies.

In Fig.~\ref{fig:deep2}, we show the cross-matched catalogue between
DEEP2 and DECaLS DR8 in red, and the TNG galaxies augmented with FSPS-synthesized
colours in blue for the three redshift samples considered 
$z = 0.8$, $z = 1.1$ and $z = 1.4$. The median redshifts for the DEEP2-selected samples are $z = 0.83$, $z = 1.08$ and $z = 1.37$, respectively. We see that the $g-r$ vs. 
$r-z$ distributions are consistent between the two samples
and are within 0.05, 0.2 and 0.1 mag for the three redshifts, respectively. These differences can be attributed to the observation that the TNG-predicted spectral energy distributions (SEDs) are fainter in the $r$-band for a given $g-z$ colour than the observed data, which yields galaxies bluer in their $g-r$ and redder in their $r-z$ bands. The $r$-band rest-frame frequencies at these redshifts correspond to the UV regime. The UV SED in the population synthesis model is sensitive to our choice for a dust attenuation model. In particular, changes in the dust law (concerning the slope as well as the UV bump) could introduce such effects, and in principle, one could experiment more with the dust attenuation model to bring better agreement between observations and simulations, but the insight would be minimal. Another possibility is related to the variability of star formation: since the UV (and in particular the far-UV) traces the SFR on short timescales \citep[e.g.,][]{2019MNRAS.487.3845C}, the UV luminosity might be underpredicted in TNG because of too smooth (not bursty enough) SFHs \citep{2015MNRAS.447.3548S,2020MNRAS.498..430I}.
We also note an overall shift of the SEDs towards the lower left corner, as we increase the samples redshift.
In addition, the photometric scatter is
quite similar as can be seen in the width of the red and blue curves
in Fig.~\ref{fig:deep2}, which validates the our heuristic approach to adding observational realism to the TNG sample.

\subsection{sSFR-selected sample}
\label{sec:meth.sfr}
Many of the current and future cosmological surveys will 
target star-forming ELGs whose spectrum is characterized 
by prominent [O II] and [O III] emission lines
as well as other less prominent features such as [Ne III] and Fe 
II$^\ast$ emission lines. These surveys will apply
a combination of colour-colour cuts as well as magnitude
cuts (see Table \ref{tab:obs} for the
particular selection choices applied to the DESI and
eBOSS ELG samples). A direct output of hydrodynamical 
simulations such as IllustrisTNG is the SFR,
which is available at every snapshot for each galaxy. 
The ELG target selection in colour-colour space aims to isolate galaxies with vigorous star formation. Therefore, it is important
to validate that indeed the selected objects have a significant
overlap with the ``true'' star-forming objects as found in the
simulation. In the following sections, we compare the colour-selected
ELG samples with samples of galaxies selected by their specific star-formation rate (i.e., the SFR per stellar mass).
In addition
to satisfying a sSFR limit, we also require of the star-forming sample 
to be within the chosen magnitude cut for the ELGs in the survey.
This corresponds to $21.825 < g < 22.825$ and $20.0 < g < 23.6$ for eBOSS and DESI, respectively. {The sSFR threshold is selected so that the number of galaxies in the sSFR-selected sample matches that of the colour-selected one and is equal to $\log[{\rm sSFR}] = -9.37$, $\log[{\rm sSFR}] = -9.23$ and $\log[{\rm sSFR}] = -9.23$  for the DESI sample at $z = 0.8$, $z = 1.1$ and $z = 1.4$, and $\log[{\rm sSFR}] = -9.69$ for the eBOSS sample at $z = 0.8$. The selection criteria for the sSFR-selected sample are summarized in Table~\ref{tab:sfr}}. We next study and compare the occupation function, two-point clustering, bias and cross-correlation coefficient of both samples.

\begin{table}
\caption{{Selection criteria for constructing a sSFR-selected sample in TNG300 that attempts to identify ELG-like galaxies at $z = 0.8$, $z = 1.1$ and $z = 1.4$ for the DESI and eBOSS galaxy surveys.}}
\begin{center}
\begin{tabular}{c|c|c|c}
\hline
Survey & Redshift & sSFR limit & Magnitude cut\\[0.1cm]
\hline \hline
eBOSS & $z=0.8$ & $\log[{\rm sSFR}] > -9.69$ &  $21.825< g < 22.825$\\[0.1cm]
\hline 
DESI  & $z=0.8$ & $\log[{\rm sSFR}] > -9.37$ &  $20.0< g < 23.6$ \\[0.1cm]
\hline 
DESI  & $z=1.1$ & $\log[{\rm sSFR}] > -9.23$ & $20.0< g < 23.6$ \\[0.1cm]
\hline 
DESI  & $z=1.4$ & $\log[{\rm sSFR}] > -9.23$ & $20.0< g < 23.6$ \\[0.1cm]
\hline
\end{tabular}
\end{center}
\label{tab:sfr}
\end{table}

\section{Results}
\label{sec:res}
\begin{table}
\caption{Expected number density ($n_{\rm gal}$, target number density) and actual number density of the simulated galaxies (achieved number density) for the eBOSS and DESI ELG samples considered in his work at $z = 0.8$, $z = 1.1$, $z = 1.4$. The galaxies are obtained after applying the colour cuts from Table~\ref{tab:obs} to the colour-synthesized galaxies in the TNG300 simulation box of volume $V_{\rm box} = 205^3 [{\rm Mpc}/h]^3$. The galaxy number densities quoted for the two surveys can be found in \citet{2016arXiv161100036D} and \citet{2017MNRAS.471.3955R} and are generally in good agreement with the derived galaxy samples after applying their respective colour selection criteria.}
\begin{center}
\begin{tabular}{c|c|c|c}
\hline
Survey & Redshift & Target number density & Achieved number density\\[0.1cm]
\hline \hline
eBOSS & $z=0.8$ & $6 \times10^{-4} \,[{\rm Mpc}/h]^{-3}$ & $9.4 \times10^{-4} \,[{\rm Mpc}/h]^{-3}$ \\[0.1cm]
\hline 
DESI  & $z=0.8$ & $9 \times10^{-4} \,[{\rm Mpc}/h]^{-3}$ & $10.0 \times10^{-4} \,[{\rm Mpc}/h]^{-3}$ \\[0.1cm]
\hline 
DESI  & $z=1.1$ & $4 \times10^{-4} \,[{\rm Mpc}/h]^{-3}$ & $5.4 \times10^{-4} \,[{\rm Mpc}/h]^{-3}$ \\[0.1cm]
\hline 
DESI  & $z=1.4$ & $1 \times10^{-4} \,[{\rm Mpc}/h]^{-3}$ & $1.8 \times10^{-4} \,[{\rm Mpc}/h]^{-3}$ \\[0.1cm]
\hline
\end{tabular}
\end{center}
\label{tab:num}
\end{table}
Here we present an analysis of the ELG-like galaxies
extracted from TNG300 using the methods described in the
previous section. In particular, we study the ELG populations
derived after applying the cuts in colour space proposed
by the DESI survey at $z = 0.8$, $z = 1.1$ and $z = 1.4$ and
by eBOSS at $z = 0.8$. We also make use of  
the sSFR-selected sample, which combines both a sSFR
cut and a magnitude cut matching that of the respective
survey (see Section \ref{sec:meth.sfr} {and Table~\ref{tab:sfr}}). First, we study the colour-(s)SFR distribution of the
TNG300 galaxies to test the robustness of the DESI cuts.
Next, we demonstrate the two-dimensional distribution of the
ELGs in the cosmic web, comparing it with that of the most
massive galaxies at the same number density. Importantly,
we also study the galaxy occupations of the haloes hosting
ELGs and
their dependence on the local halo environment. Finally, we
study the clustering properties of the samples, focusing on their dependence on environment, as well
as their galaxy bias and cross-correlation coefficient.

\subsection{Number densities of ELGs}

The expected number densities of ELGs at various redshifts
for DESI and eBOSS can be found in \citet{2016arXiv161100036D} and \citet{2017MNRAS.471.3955R}.
For the three redshifts we consider ($z = 0.8$, $z = 1.1$, and $z = 1.4$), the number densities of ELGs for the DESI
survey are $n_{\rm gal} \approx 9 \times10^{-4} \,[{\rm 
Mpc}/h]^{-3}$, $n_{\rm gal} \approx 4 \times10^{-4} \,[{\rm 
Mpc}/h]^{-3}$ and $n_{\rm gal} \approx 1 \times10^{-4} \,[{\rm 
Mpc}/h]^{-3}$, respectively, and for the eBOSS ELGs at $z = 0.8$, it is $n_{\rm gal} \approx 6 \times10^{-4}
\,[{\rm Mpc}/h]^{-3}$. After applying the cuts in Table
\ref{tab:obs} on the synthesized galaxies
in our hydro simulation box TNG300 ($V_{\rm box} = 205^3
[{\rm Mpc}/h]^3$), we find 8637  ($n_{\rm gal} \approx 10.0 \times 10^{-4} \,[{\rm Mpc}/h]^{-3}$), 4632 ($n_{\rm gal} \approx 5.4 \times10^{-4} \,[{\rm Mpc}/h]^{-3}$), and 1539 ($n_{\rm gal} \approx 1.8 \times10^{-4} \,[{\rm Mpc}/h]^{-3}$) DESI ELG-like
objects at $z = 0.8$, $z = 1.1$, and $z = 1.4$, respectively,
and 8112 ($n_{\rm gal} \approx 9.4 \times10^{-4} \,[{\rm Mpc}/h]^{-3}$) eBOSS ELGs at $z = 0.8$. These are in
good agreement with the expected number densities of the surveys
stated above. These numbers are also shown in Table \ref{tab:num}
for clarity.

\subsection{colour-(s)SFR dependence}
\label{sec:res.col-sfr}
An intriguing question to pursue is what the distribution
of the TNG300 galaxies in SFR-colour and sSFR-colour space looks like
in relation to the ELG cuts performed by 
galaxy survey analyses. 
Here, sSFR is the specific SFR, which is defined
as the SFR per stellar mass, which 
roughly probes the inverse of the doubling timescale of the stellar mass
growth due to star formation \citep{2007ApJ...660L..43N}. In star-forming
galaxies, as redshift decreases, ongoing star formation increases stellar mass, while the sSFR typically declines.
{Because galaxy surveys targeting
ELGs such as DESI and eBOSS perform their selection
with the goal of identifying luminous star-forming ELGs, their eligibility
criteria (beyond the magnitude cut of Table \ref{tab:obs})
are more akin to a cut in sSFR rather than SFR.}

\begin{figure}
\centering
\includegraphics[width=1.\columnwidth]{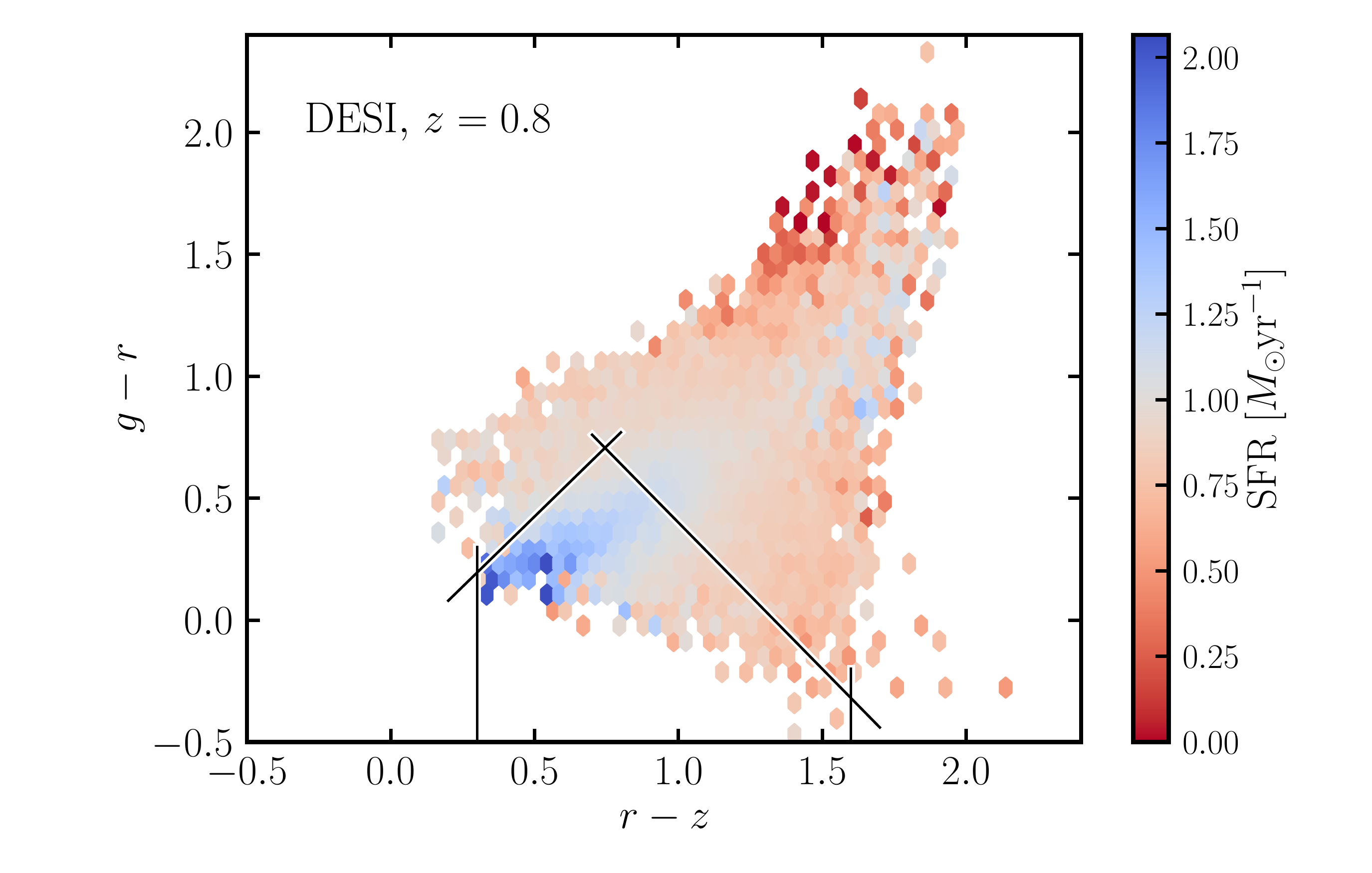} \\
\includegraphics[width=1.\columnwidth]{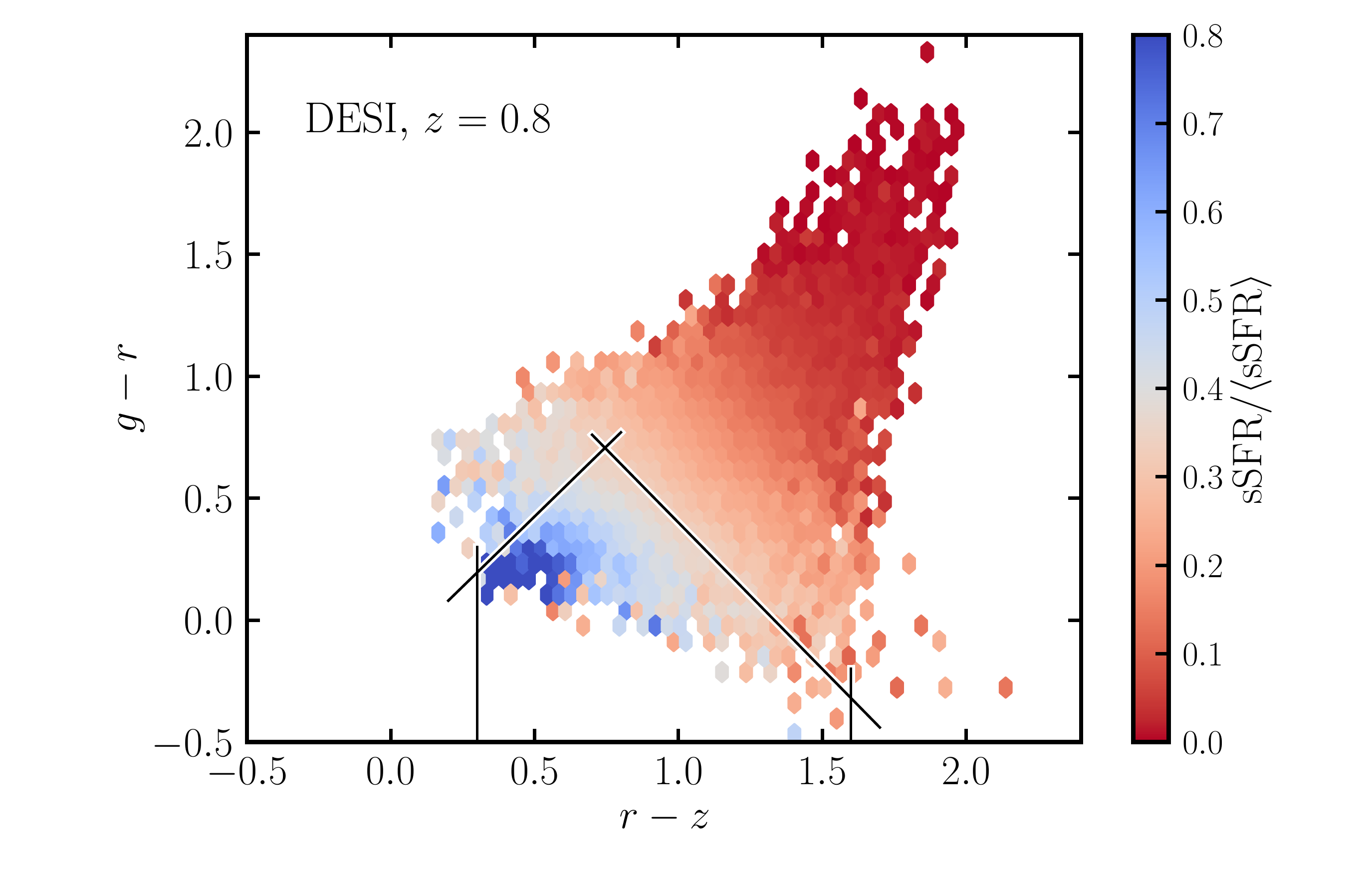}
\caption{Representation of the TNG300 galaxies at $z = 0.8$ in colour-colour space. The hexagonal shapes are coloured based on star-formation rate (SFR, \textit{top panel}) and specific star-formation rate (sSFR, \textit{bottom panel}) with higher values corresponding to bluer colours and lower values corresponding to redder colours. We have applied the DESI magnitude limit and $g$-band cuts from Table~\ref{tab:obs} to the synthetic sample. In \textit{black}, we show the colour selection cuts proposed by the DESI collaboration for selecting emission-line galaxies (ELGs). {We can see that high SFR values are found in both the blue and red loci of the diagram, whereas galaxies with high sSFR values are highly concentrated in the region of the blue young and less massive galaxies. The colour-colour DESI cut aims to select precisely yougn star-forming objects and is therefore most closely mimicked by a cut in sSFR (accompanied by the magnitude cut given in Table \ref{tab:obs}).}}
\label{fig:selection}
\end{figure}

In Fig.~\ref{fig:selection}, we visualize the SFR- and 
sSFR-colour three-dimensional diagrams for the TNG300 galaxies
at $z = 0.8$ alongside the DESI colour-colour cuts. The
\textit{top panel} demonstrates that the SFR values
are much more evenly distributed across the galaxy population
compared with the sSFR ones, as expected. There are two
modes corresponding to the rapidly star-forming young galaxies,
which are captured well by the DESI cuts, and the very massive
red galaxies that also undergo significant amounts of star formation.
However, as can be seen in the \textit{bottom panel}, their
SFRs are negligible compared with their masses, so the sSFRs
are very low for these objects. {Both the sSFR and the DESI cuts thus successfully isolate young vigorously
star-forming galaxies, as intended.}

\subsection{Spatial distribution of ELGs}
\label{sec:res.2d}

A large number of observational studies such as GAMA
\citep{2018MNRAS.474..547K},
VIPERS \citep{2017MNRAS.465.3817M}, and
COSMOS \citep{2018MNRAS.474.5437L} have found that
star-forming and less massive galaxies 
are more likely to reside in filamentary regions  than 
quiescent and more massive  galaxies. Filaments are also
believed to assist gas cooling, thus
enhancing star-formation processes in galaxies
\citep{2020ApJ...899...98V}. Similar results have been found
in hydrodynamical simulations \citep[e.g.][]{2019MNRAS.485..464L}.  
These results can be explained as a combination of
two effects: accretion in filaments is predominantly
smooth \citep[e.g.][]{2019MNRAS.483.3227K}, and their outskirts
are vorticity rich \citep{2015MNRAS.446.2744L}. Furthermore,
the majority of the cold gas, which is needed for star formation
to take place, is located in filaments and sheets
at $z \sim 1$, which provides insight into why
ELGs and star-forming galaxies are overwhelmingly found in these
regions \citep{2012MNRAS.423.2279C,2014MNRAS.437..816C,2019MNRAS.485.2367C}.

Conventionally, the cosmic web is classified into  voids,
sheets, filaments and knots \citep{1989Sci...246..897G}.  To characterize the
the large-scale distribution of our simulation, 
we follow the traditional characterization
into tidal environments
\citep{1970Ap......6..320D,2007MNRAS.375..489H,2009MNRAS.396.1815F}.
First, we evaluate the discrete density field,
$\delta (\mathbf{x})$, using cloud-in-cell (CIC) interpolation on a 
$128^3$ cubic lattice of all particles in TNG300-3-Dark\footnote{See \href{www.tng-project.org} for details on
this simulation}. We then apply a Gaussian smoothing kernel
with a smoothing scale of $R_{\rm smooth} = 4 \ {\rm Mpc}/h$.
{For simplicity, we work with a rescaled version of 
the Newtonian potential, $\phi \equiv \bar \phi /(4 \pi G 
\bar \rho)$,  for  which  the  Poisson  equation  is  
simply $\nabla^2 \phi = \delta$, where $\delta$ is the 
matter overdensity field. We then define  the  tidal  
tensor  field  as $t_{ij} \equiv \partial_i \partial_j \phi$, and  we classify the environment according to the number of  eigenvalues, $\lambda_1 \geq \lambda_2 \geq \lambda_3$, above a given threshold, $\lambda_{\rm th}$:}
\begin{itemize}
\item \textbf{peaks}: all eigenvalues below the threshold ($\lambda_{\rm th} \geq \lambda_1$)
\item \textbf{filaments}:  one eigenvalue below the threshold ($\lambda_1 \geq \lambda_{\rm th} \geq \lambda_2$)
    \item \textbf{sheets}:  two eigenvalues below the threshold ($\lambda_2 \geq \lambda_{\rm th} \geq \lambda_3$)
    \item \textbf{voids}: all eigenvalues above the threshold ($\lambda_3 \geq \lambda_{\rm th}$).
\end{itemize}
{This formalism has one free parameter: the eigenvalue threshold $\lambda_{\rm th}$, and there have been several  prescriptions proposed in the literature to choose a value for it. A choice of $\lambda_{\rm th} = 0$ would separate different regions based purely on the direction of the tidal forces. However, this prescription would assume that gravitational  collapse  is  underway  along  a  given  direction even  if  the  eigenvalue  is  only  infinitesimally  positive and collapse would only occur after a very long time. This approach thus produces a tidal classification in which voids occupy only about $\sim$20\% of the volume, in striking contrast with the visual impression from redshift surveys and $N$-body simulations that most of the volume is actually  empty. On the other hand, a choice of $\lambda_{\rm th}>0$ only regards a given direction  as  ``collapsing''  if  the  tidal forces are sufficiently strong and should give rise to a tidal classification in which the abundance of voids better matches our intuitive expectations. To choose a value for $\lambda_{\rm th}$, we follow a procedure  similar  to \citet{2015MNRAS.448.3665E} and \citet{2016MNRAS.460..256A}:  for  different  values  of  the  eigenvalue  threshold  we calculate  the  number  of  galaxies  in  the  simulation located  in the three different environmental types, and we choose the value of $\lambda_{\rm th}$ that most equally divides the galaxy population among  the  different  types.}
We pick $\lambda_{\rm th} = 0.1$, which yields
a balanced distribution that agrees qualitatively with a rough visual classification into the four tidal field components.

\begin{figure}
\centering
\includegraphics[width=1.\columnwidth]{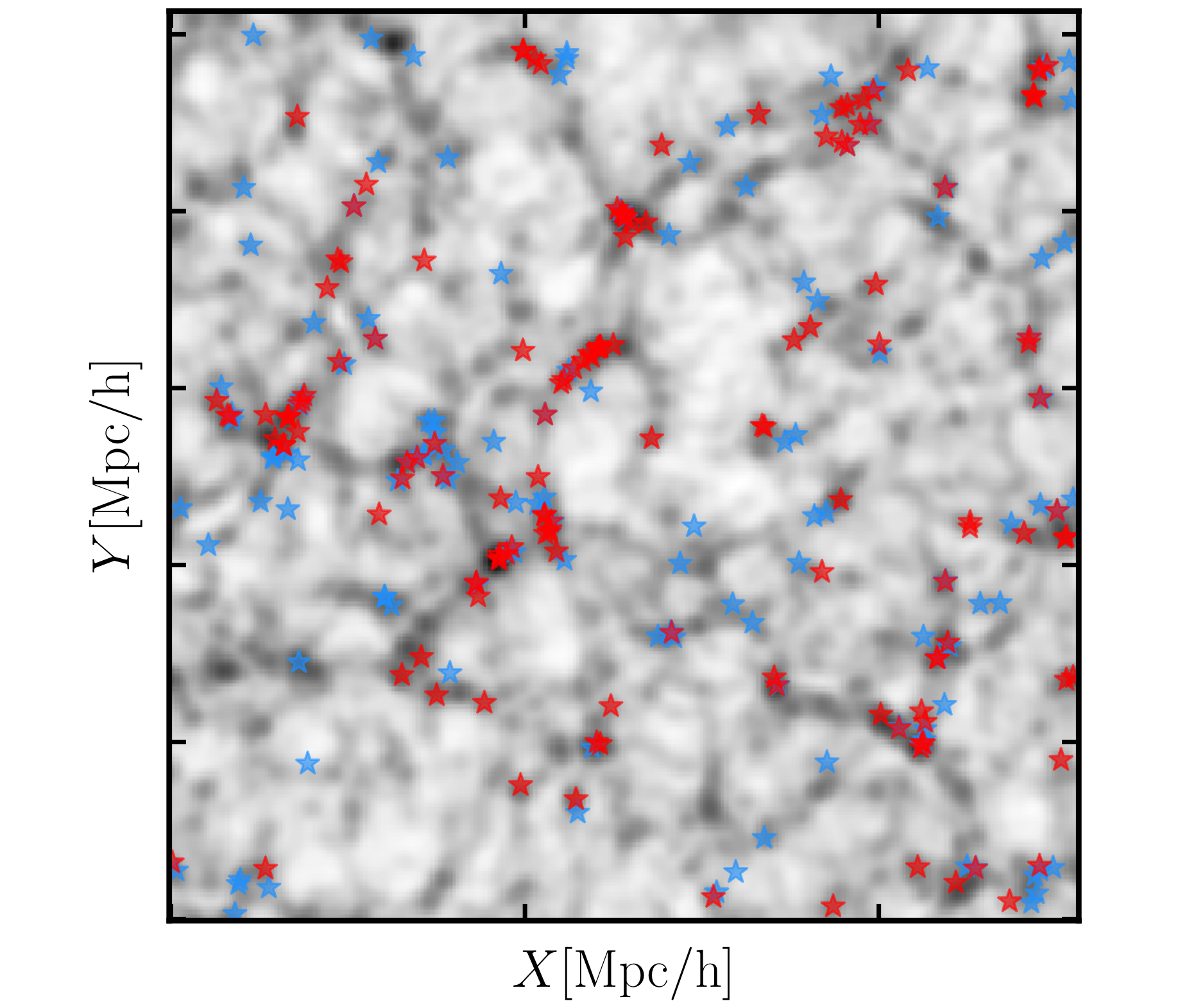} \\
\includegraphics[width=1.\columnwidth]{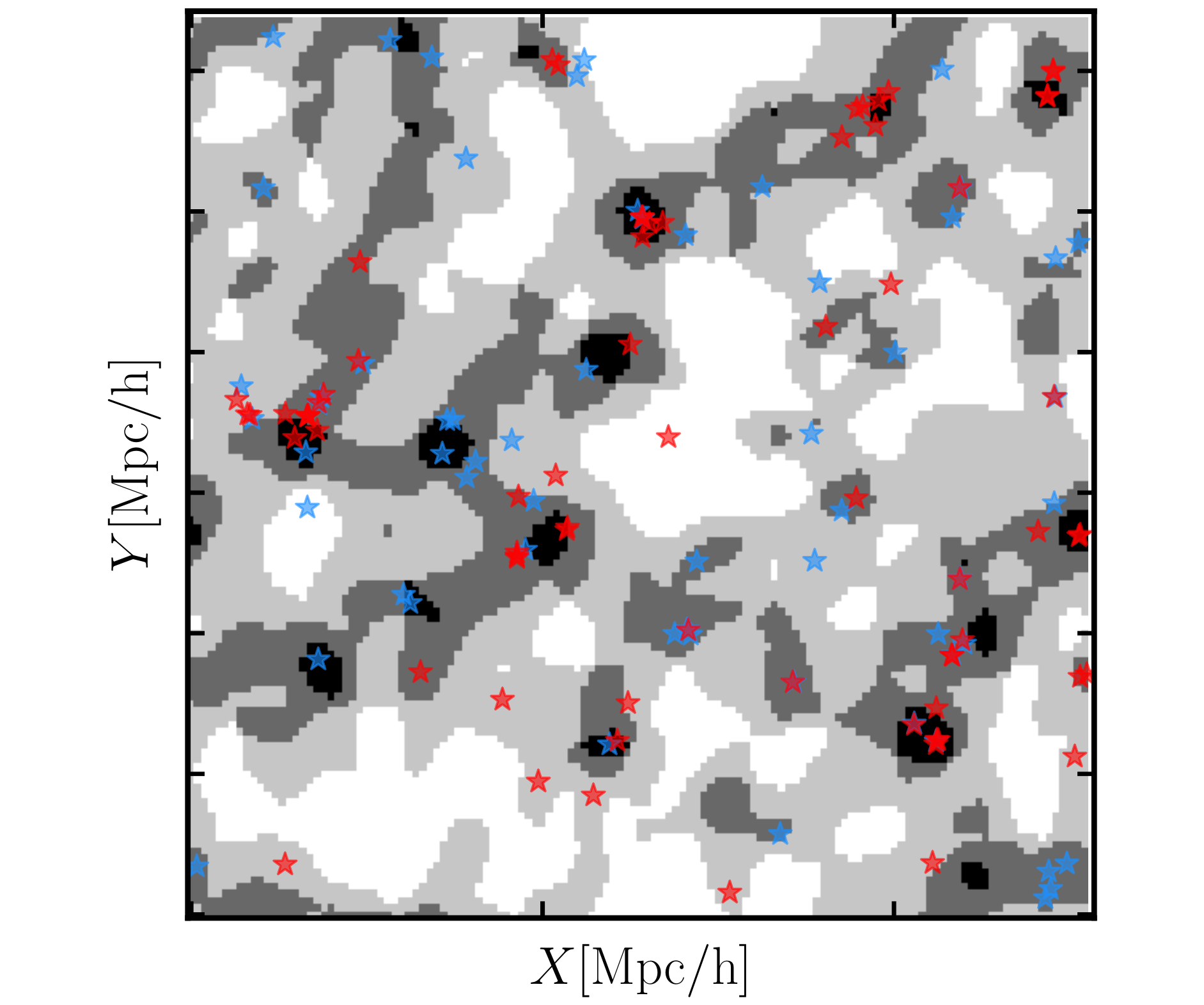} \\
\includegraphics[width=1.\columnwidth]{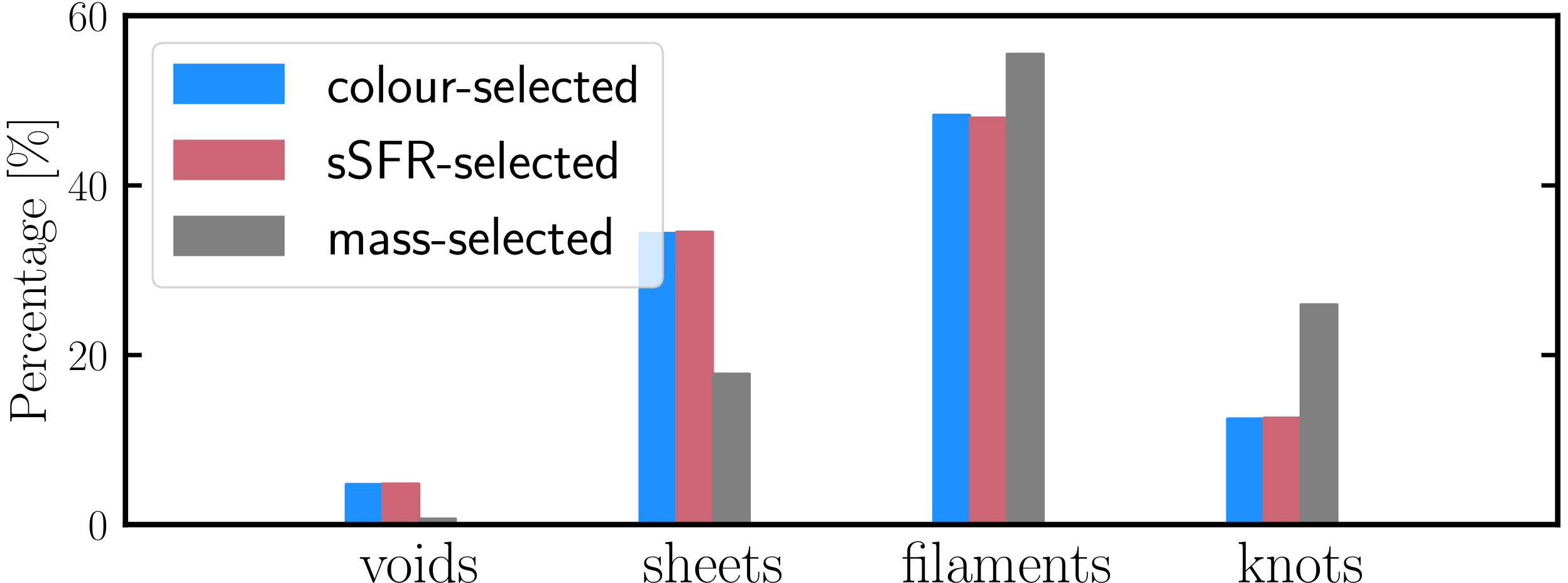}
\caption{Galaxy distribution of DESI colour-selected
ELGs (in \textit{blue}) and stellar-mass-selected galaxies
(in \textit{red}) at $z = 0.8$ with a  number  density of 
$n_{\rm gal} = 1.0 \times10^{-3} \,[{\rm Mpc}/h]^{-3}$,
shown as a horizontal cross section of the TNG300 box ($L_{\rm box} = 205 \ {\rm Mpc}/h$) of depth $\sim~3 \ {\rm Mpc}/h$. On the
\textit{top panel}, the galaxies are displayed on top of
the smoothed dark matter density field (with a Gaussian smoothing
scale of $R_{\rm smooth} = 1.5 \,{\rm Mpc}/h$)
in \textit{grey}. The \textit{middle panel} shows the galaxies painted
over the four traditional tidal environment types -- 
knots in \textit{black}, filaments in \textit{dark grey}, sheets in
\textit{light grey}, and voids in \textit{white}, as defined in
Section \ref{sec:res.2d}. This slice has a thickness of $1.6 \,{\rm 
Mpc}/h$. We see that the ELG sample of star-forming galaxies
is distributed predominantly in filamentary structures, while
the stellar-mass selected galaxies are mostly found in the highest
density regions, i.e. the knots. Across the entire volume of the simulation, for the colour-selected sample, we find that 4.8\%, 34.4\%, 48.3\%, and 12.5\% of the galaxies live in voids, sheets, filaments, and knots, respectively, whereas the analogous percentages for the mass-selected sample are 0.7\%, 17.8\%, 55.5\%, and 26.0\%. {The fraction of galaxies living in each environment type is shown in the \textit{bottom panel}.} This clearly demonstrates that ELGs have a stronger preference to inhabit the less dense regions of the simulation.}
\label{fig:2d_distn}
\end{figure}

In Fig.~\ref{fig:2d_distn}, we show the two-dimensional distribution
of the ELG galaxy sample along with the mass-selected one. The ELG
sample displayed in the figure corresponds to the colour-selected
galaxies at $z = 0.8$, following the DESI $g-r$ and $r-z$ cuts.
The stellar-mass cut has been made so that the number density of
the objects in both samples is the same, at
$n_{\rm gal} = 9 \times10^{-4} \,[{\rm Mpc}/h]^{-3}$ (see Table \ref{tab:num}). The stellar-mass selected sample serves as a proxy for an LRG-like sample.
In agreement with previous studies, most of the galaxies in the
star-forming and mass-selected samples populate filaments. 
The mass-selected sample is roughly twice as likely to be present
in knots and half as likely to be found in sheets ({see the bottom pannel of Fig.~\ref{fig:2d_distn}}).
Across the entire volume of the simulation, we find that 4.8\%,
34.4\%, 48.3\%, and 12.5\% of the colour-selected galaxies live 
in voids, sheets, filaments, and knots, respectively, whereas 
the corresponding percentages for the mass-selected sample are
0.7\%, 17.8\%, 55.5\%, and 26.0\%. This clearly demonstrates 
that ELGs have a stronger preference to occupy the less dense
regions of the simulation. We also observed that choosing a lower 
bound on the galaxy number density, i.e. including fewer galaxies,
leads to a more pronounced presence in knots in both the 
mass-selected and ELG samples.

\subsection{The Halo Occupation Distribution of ELGs}
\label{sec:res.hod}
The HOD represents a useful approach for the construction
of mock catalogues because it assumes a simple and
readily implementable relation
between halo mass and occupation number. It is also
one of the fundamental statistics used to study the
galaxy-halo connection, as it provides information about
the average number of galaxies as a function of halo mass.

\begin{figure*}
\centering
\includegraphics[width=0.48\textwidth]{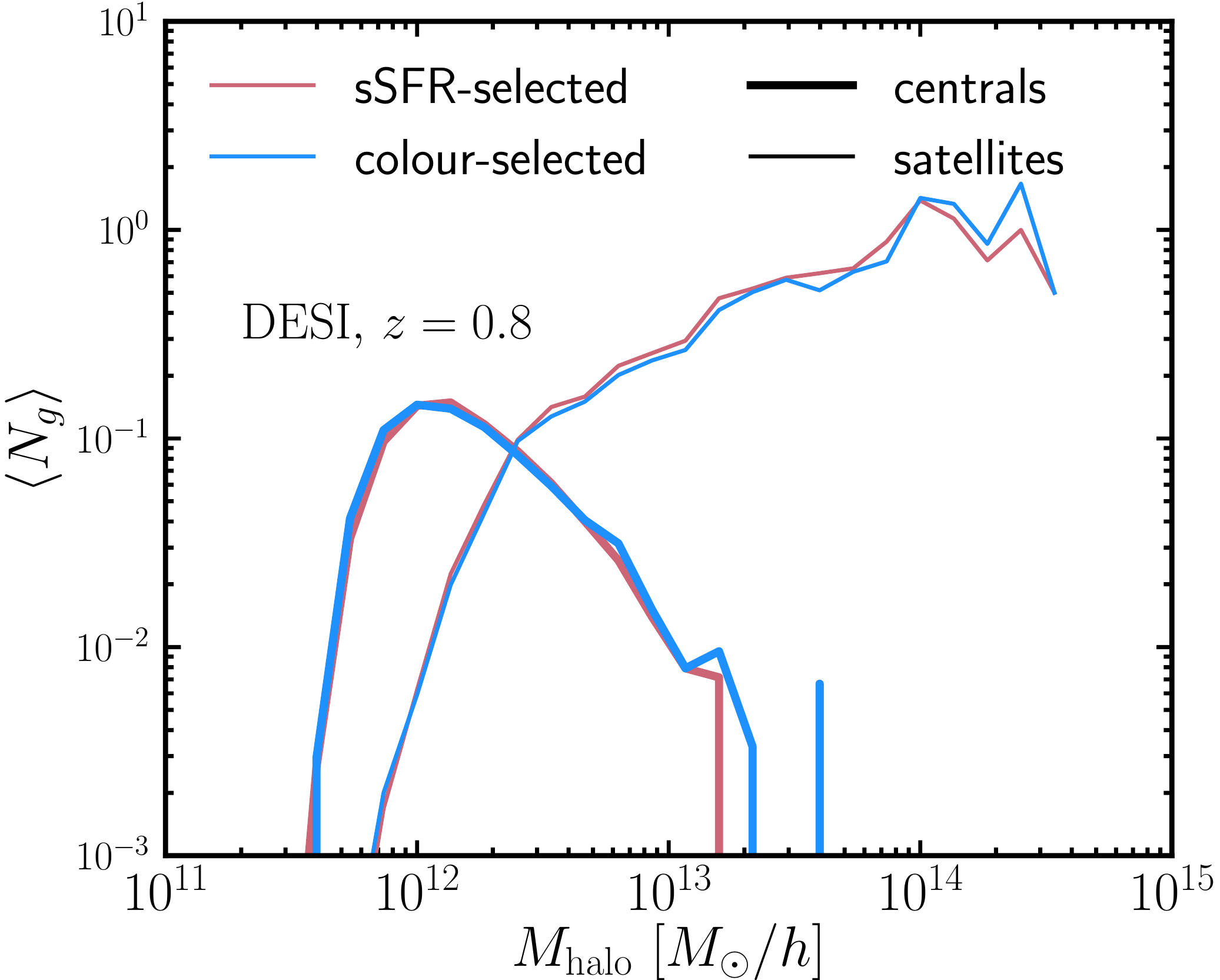}
\includegraphics[width=0.48\textwidth]{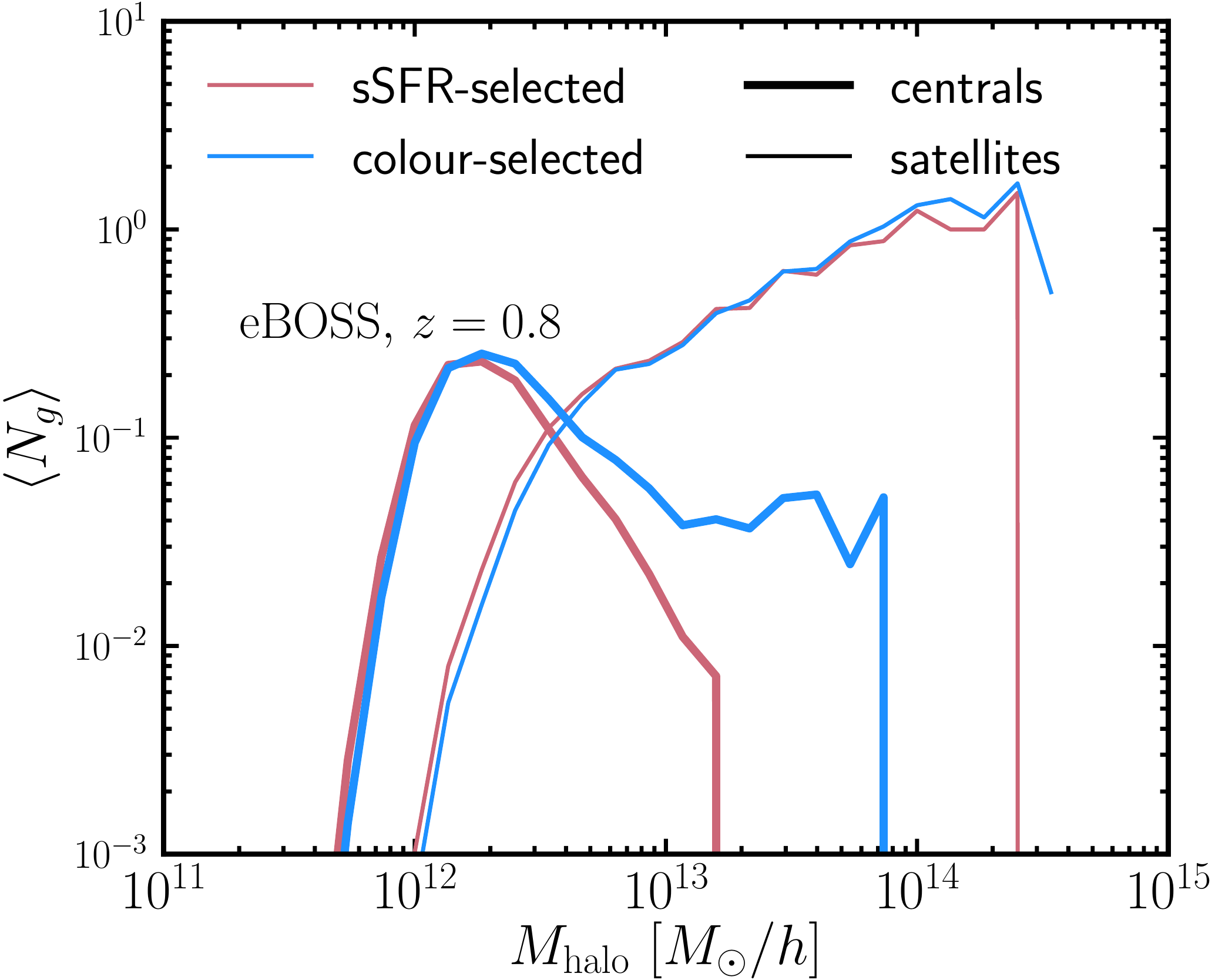} \\
\includegraphics[width=0.48\textwidth]{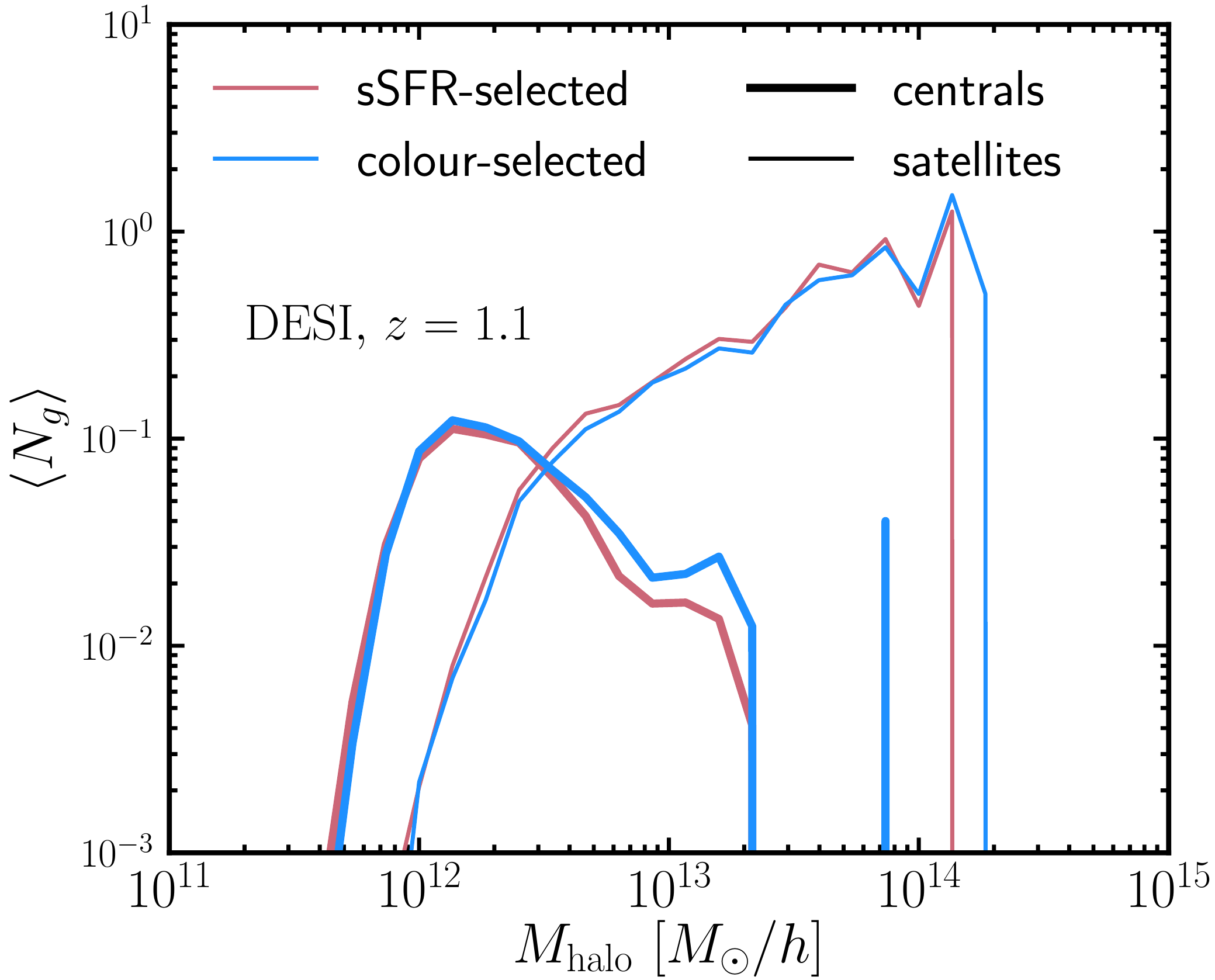}
\includegraphics[width=0.48\textwidth]{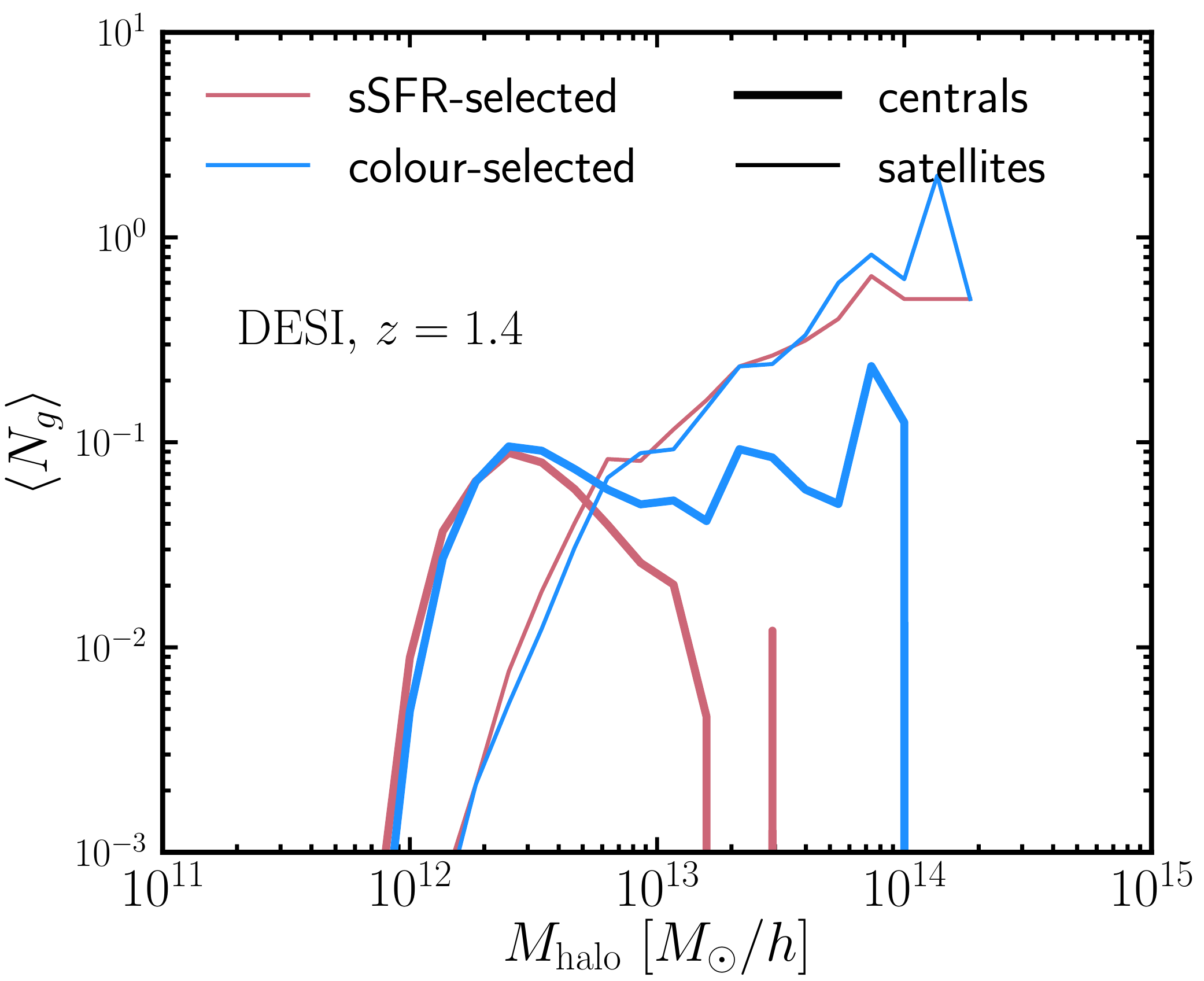}
\caption{HODs of the colour- and sSFR-selected ELGs in TNG300 split into centrals (\textit{thick lines}) and satellites (\textit{thin lines}). The colour selection  (in \textit{blue}) is done using the DESI and eBOSS cuts (see Table \ref{tab:obs}), while the sSFR selection (in \textit{red}) is detailed in Section \ref{sec:meth.sfr} and also attempts to target ELGs. In the \textit{top} plots, we demonstrate the DESI- (\textit{left}) and eBOSS- (\textit{right}) selected ELG samples at $z = 0.8$, and on the \textit{bottom}, we show the DESI-selected samples at $z = 1.1$ (\textit{left}) and $z = 1.4$ (\textit{right}). We see that the agreement between the two selection strategies is excellent for the DESI $z = 0.8$ sample and gets worse with redshift (see the $z = 1.1$ and $z = 1.4$ panels) for the HOD of the centrals. This is likely the case since at higher redshifts, the colour-colour distribution of galaxies shifts towards the DESI ELG-targeting boundaries, and a large number of red galaxies with low sSFR are selected (see Fig.~\ref{fig:deep2}). This results in an extra contribution to the HOD of centrals, hosted by more massive haloes. The case for the eBOSS sample is similar, for which the satellite population is in a very good agreement, while for the centrals, the colour cuts yield a larger number of centrals at higher halo mass.}
\label{fig:hod}
\end{figure*}

The HOD is usually broken down into the contribution
from central and satellite galaxies. In Fig.~\ref{fig:hod},
we present the sSFR- and colour-selected ELG-like galaxies
for the DESI and eBOSS surveys for the three redshift
samples of interest $z = 0.8$, $z = 1.1$ and $z = 1.4$. 
For each redshift choice, the number density is the same
as that stated in Table \ref{tab:num}.
The thick and thin curves show separately the
contribution to the halo occupation from centrals
and satellites, respectively, which appears to be
distinct from that of a stellar-mass selected sample.
The HOD shape of mass-selected centrals is typically similar
to a smoothed step function, in which haloes gradually 
transition to being guaranteed to host a central galaxy 
above a certain mass threshold. Both in our ELG-like HODs
as well as in the traditional mass-selected samples, 
the satellite occupation roughly follows a power law above 
some halo mass limit.

Each HOD is computed in logarithmic mass bins of width 0.1 dex,
and the average occupation number
is plotted at the central value of each bin. The galaxies in both
our ELG-like samples correspond mainly to blue star-forming
galaxies and exclude luminous red galaxies with high stellar
mass but low sSFR. While the ranking of galaxies in order of their
emission-line luminosity is not equivalent to a sSFR ranking 
due to dust attenuation, i.e. the most rapidly star-forming
galaxies may not have the brightest emission lines, the
additional magnitude cut that we apply (see Section \ref{sec:meth.sfr} {and Table~\ref{tab:sfr}})
makes the two samples extremely congruous. This can be seen
in Fig.~\ref{fig:hod}, which demonstrates the striking similarity
between the colour- and sSFR-selected samples (blue and red
curves, respectively). The worst agreement between the two is
found for the DESI $z = 1.4$ and the eBOSS $z = 0.8$ samples. 
For the DESI sample, this is likely the case since at higher redshifts, the
colour-colour distribution of galaxies shifts towards the DESI
ELG-targeting boundaries, and a large number of red galaxies 
with relatively lower sSFRs are selected (see Fig.~\ref{fig:deep2}). Furthermore, the magnitude selection in the $g$-band is picking intrinsically more luminous objects at higher redshifts.
This results in an extra bump to the HOD of centrals for more 
massive haloes relative to the sSFR-selected sample. Studying 
the eBOSS colour-colour-(s)SFR plot (analogously to 
Fig.~\ref{fig:selection}), we found that our synthetic eBOSS
colour-selected sample includes a significant number of massive
quiescent galaxies, resulting in a larger contribution to the 
halo occupation of massive haloes relative to the sSFR-selection choices, which tend to select smaller haloes.

\begin{figure}
\centering  
\includegraphics[width=.5\textwidth]{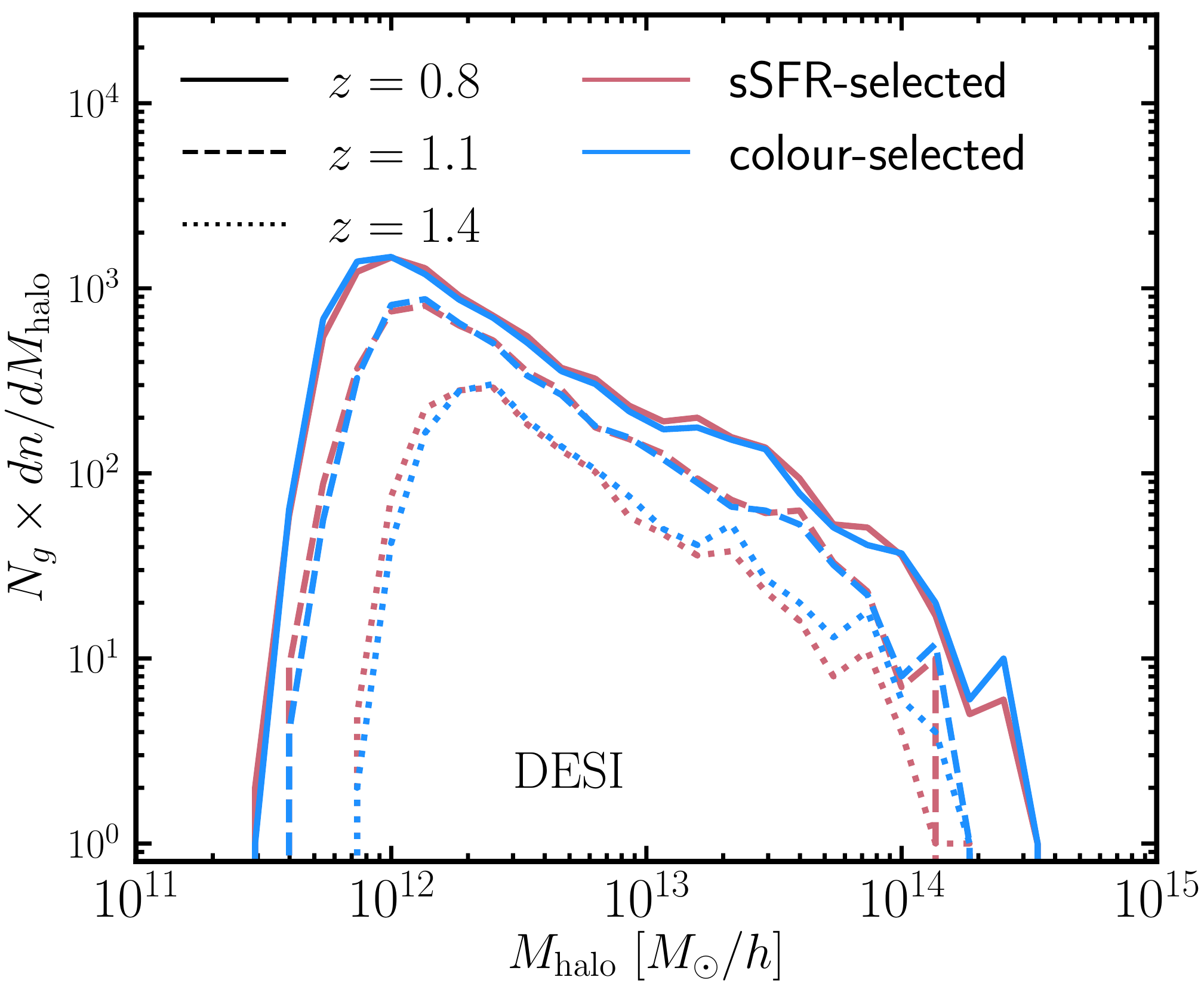}
\hfill
\caption{Cumulative occupation distributions (total number of galaxies per mass bin, $\langle N_g \rangle (dn/dM_{\rm halo})$) of the colour- and sSFR-selected galaxies in TNG300 for the redshift samples $z = 0.8$ (\textit{solid}), $z = 1.1$ (\textit{dashed}), and $z = 1.4$ (\textit{dotted}). The colour selection (in \textit{blue}) is done using the DESI cuts (see Table \ref{tab:obs}), while the sSFR selection (in \textit{red}) is detailed in Section \ref{sec:meth.sfr} and also attempts to target ELGs. We see that the agreement between the two selection strategies is very good across all redshift samples, and the largest discrepancies are visible at $z = 1.4$ (as also demonstrated in Fig.~\ref{fig:hod}). We also notice that the largest number of galaxies are found in relatively low-mass haloes ($M_{\rm halo}\sim 10^{12} \ M_\odot/h$), while the most massive haloes contribute only a modest fraction to the total number of galaxies. Furthermore, with increasing redshift, the HOD shapes are shifted to the right and the number of objects decreases gradually over all mass scales (also see Table \ref{tab:num}). For this plot, we do not differentiate between centrals and satellites, but rather display the total contribution to the cumulative occupation distribution.}
\label{fig:hod_manyz}
\end{figure}

A notable feature is that the fraction of haloes 
containing a central falls short of reaching unity in all cases
shown, suggesting that a large number of the haloes hosting ELG-like
galaxies do not actually contain a central since its emission
lines are too weak. This has also been inferred in studies involving
sSFR-selected samples in SAMs \citep[e.g.][]{2013MNRAS.432.2717C,
2019MNRAS.484.1133C,2018MNRAS.474.4024G} as well as blue galaxies in observations
\citep[e.g.][]{2011ApJ...736...59Z}. In Fig.~\ref{fig:hod_manyz}, we demonstrate
the cumulative occupation distribution of both satellites and centrals 
for the DESI colour-selected sample (in blue) and the sSFR-selected 
sample (in red) for the three redshifts $z = 0.8$, $z = 1.1$, and
$z = 1.4$. We measure this quantity as the total number of galaxies per mass bin, $\langle N_g \rangle (dn/dM_{\rm halo})$. This is a useful quantity as it clearly demonstrates that the largest number of galaxies are found in relatively low-mass haloes ($M_{\rm halo}\sim 10^{12} \ M_\odot/h$), while the most massive haloes contribute only a modest fraction to the total number of galaxies. As indicated in Fig.~\ref{fig:hod_manyz} and in Table~\ref{tab:num}, the total
number of selected objects decreases with redshift uniformly across all scales, as their detection through the DESI experiment becomes more challenging. We also see that with increasing redshift, the cumulative occupation distributions
shift to higher halo masses for all mass scales, suggesting that the
selection cuts become more sensitive to more massive ELGs. This is the case since at earlier times, massive galaxies in TNG300 are less quiescent and their emission lines are stronger, which renders them eligible for our sSFR- and colour-selection criteria.

\subsubsection{Environment dependence}
\label{sec:res.hod.env}
Studying the dependence of halo occupation on secondary properties
besides halo mass can provide us with useful insights into the
effects of assembly bias on galaxy clustering. Previous works
have studied in detail the connection between the occupation 
function and assembly bias for various halo parameters such 
as environment and concentration \citep[e.g.][]{2018ApJ...853...84Z,2019MNRAS.484.1133C}. 
\citet{sownak} perform such an analysis performed for IllustrisTNG,
finding that at fixed halo mass, there are significant
differences between the HOD shape for haloes with varying
concentration, formation epoch, and local environment. 

\begin{figure}
\centering  
\includegraphics[width=1.\columnwidth]{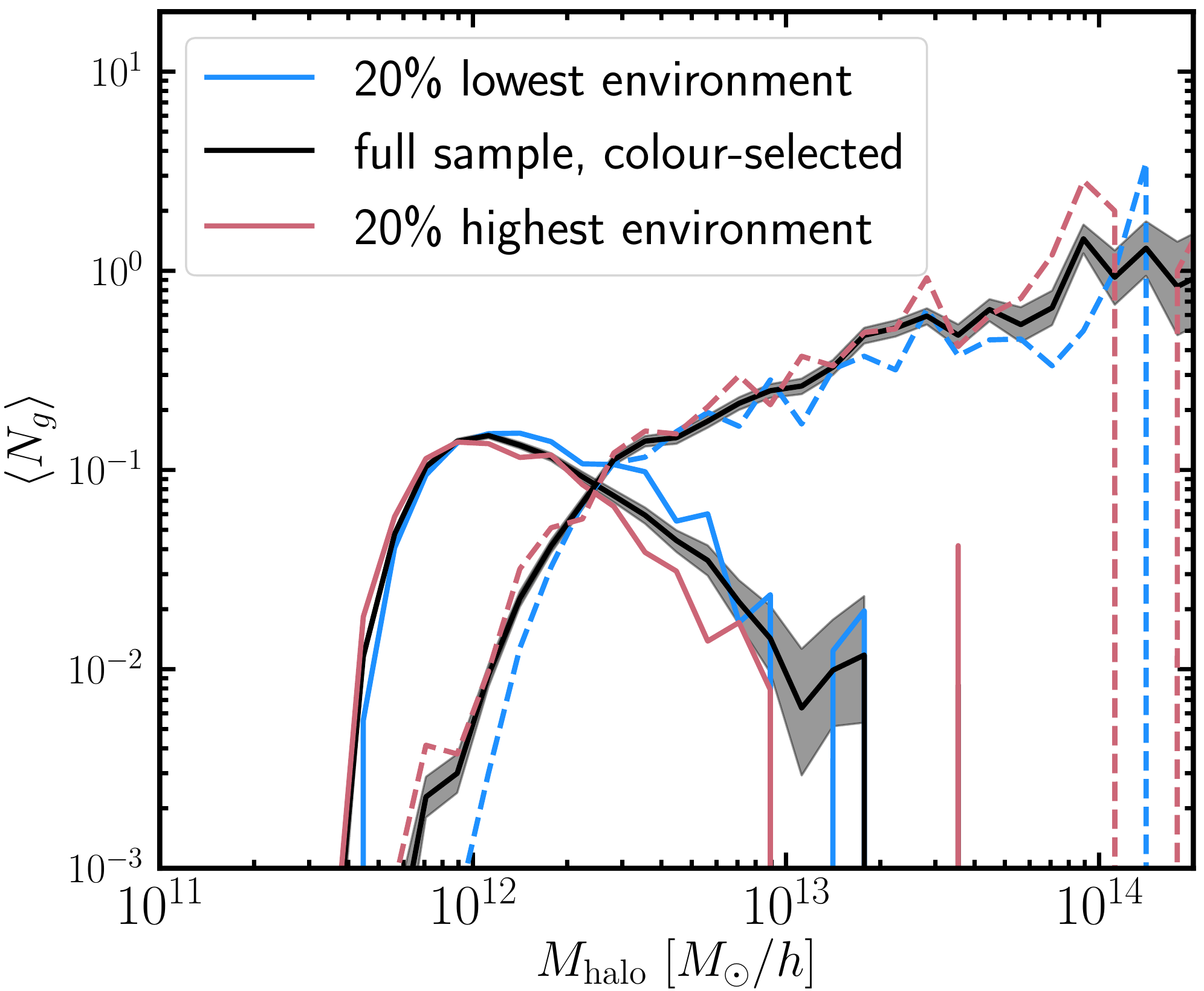} \\
\includegraphics[width=1.\columnwidth]{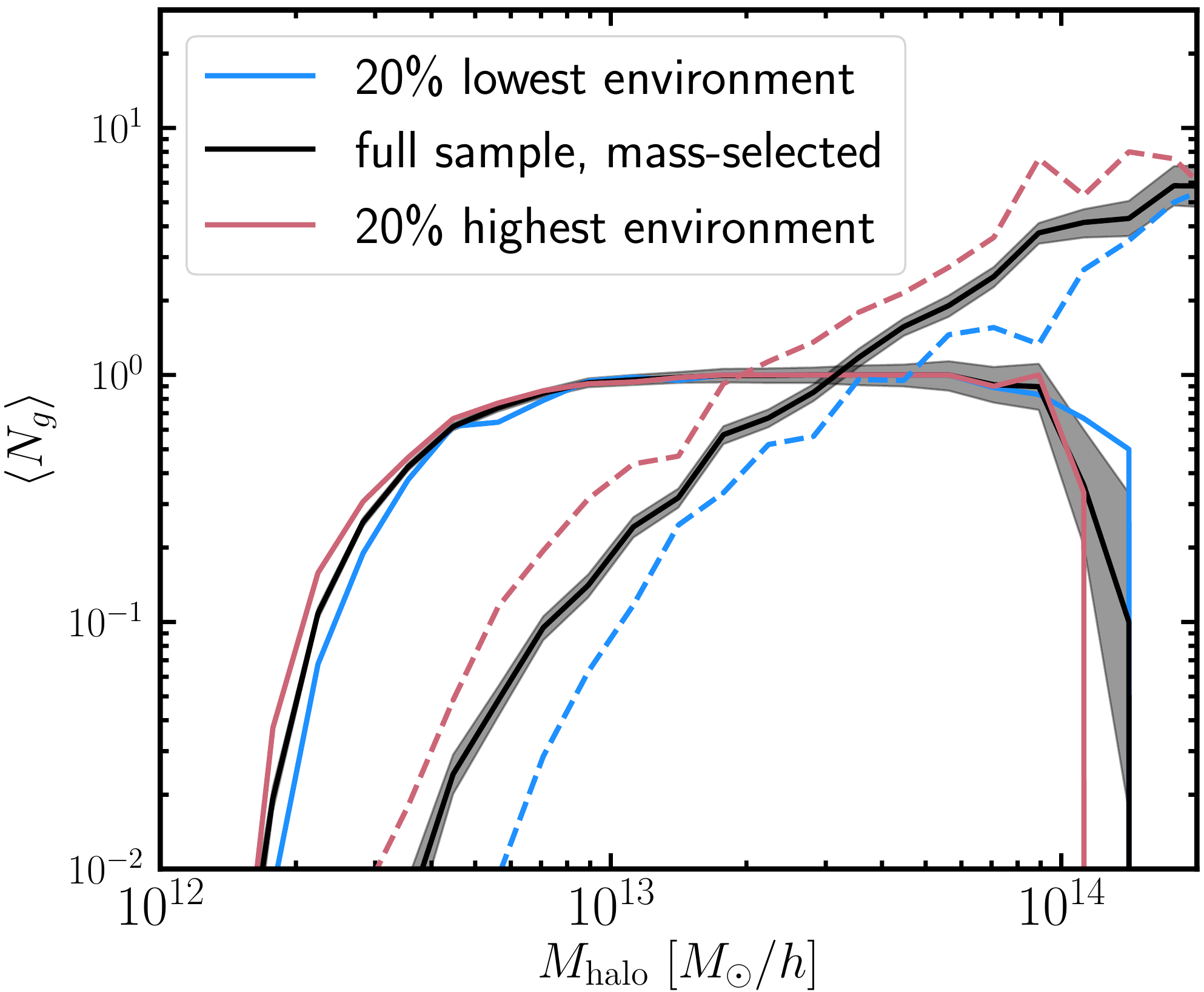}
\caption{HODs for both centrals (\textit{solid}) and satellites (\textit{dashed}) of the TNG300 galaxies split into objects living in the 20\% highest (in \textit{red}) and lowest (in \textit{blue}) environments in each mass bin. The colour-selected sample on the \textit{top panel} is obtained by applying the DESI cuts (see Table \ref{tab:obs}) at $z = 0.8$, while the \textit{bottom panel} demonstrates the environment dependence of the mass-selected sample matching the number density of the DESI ELGs (stated in Table \ref{tab:num}). {In dark grey, we show the Poisson errors.} We see that for both the qualitative dependence on environment is similar. High density environments are more conducive to hosting a larger number of galaxies, and this dependence appears to be stronger for the mass-selected sample. The ELG centrals show a slight predilection for residing in lower density environments at the high halo mass end ($\sim~10^{13}$). We also note that the overall shape of the HOD differs significantly between the two samples for the central galaxies: in the case of the mass-selected sample, it saturates at one, whereas for the ELG sample, it barely reaches 20\%.}
\label{fig:hod_env}
\end{figure}

In Fig.~\ref{fig:hod_env}, we show what the environment
dependence looks like for the colour-selected sample of
ELG-like galaxies at redshift $z = 0.8$ (\textit{top panel})
obtained after applying the DESI cuts (see Table
\ref{tab:obs}) as well as for the stellar-mass-selected sample
(\textit{bottom panel}) at the same number density
(stated in Table \ref{tab:num}). We define halo environment here as
the Gaussian-smoothed matter density over a scale of $R_{\rm smooth}
= 1.4 \ {\rm Mpc}/h$. The smoothing scale is chosen to be in the
transition regime between the one- and two-halo terms, making it
sensitive to cluster dynamics (e.g. accretion and merger events) on the outskirts of large haloes. The black
curves in both panels show the HOD for the full sample, while
the blue and red curves correspond to the average occupation
numbers of haloes living in the 20\% highest and 20\% lowest density
environments, respectively. In both cases, we see that the environment
plays a significant role in determining the number of galaxies
contained in a halo, although the effects are more apparent 
for the mass-selected case and in particular, for the satellite
contribution to the HOD. For the ELG sample, the effect is more
pronounced for haloes of lower mass, although their numbers are 
much more limited. Almost uniformly, haloes living in denser
environments tend to host a larger number of both central and
satellite ELGs. It is intriguing to note that there is a slight trend indicating
an inversion of that relation for the higher mass halo hosts,
so that low-density environments seem to
be marginally more conducive to hosting central ELGs. 

\subsection{The clustering of ELGs}
\label{sec:res.clust}

\begin{figure*}
\centering
\includegraphics[width=.48\textwidth]{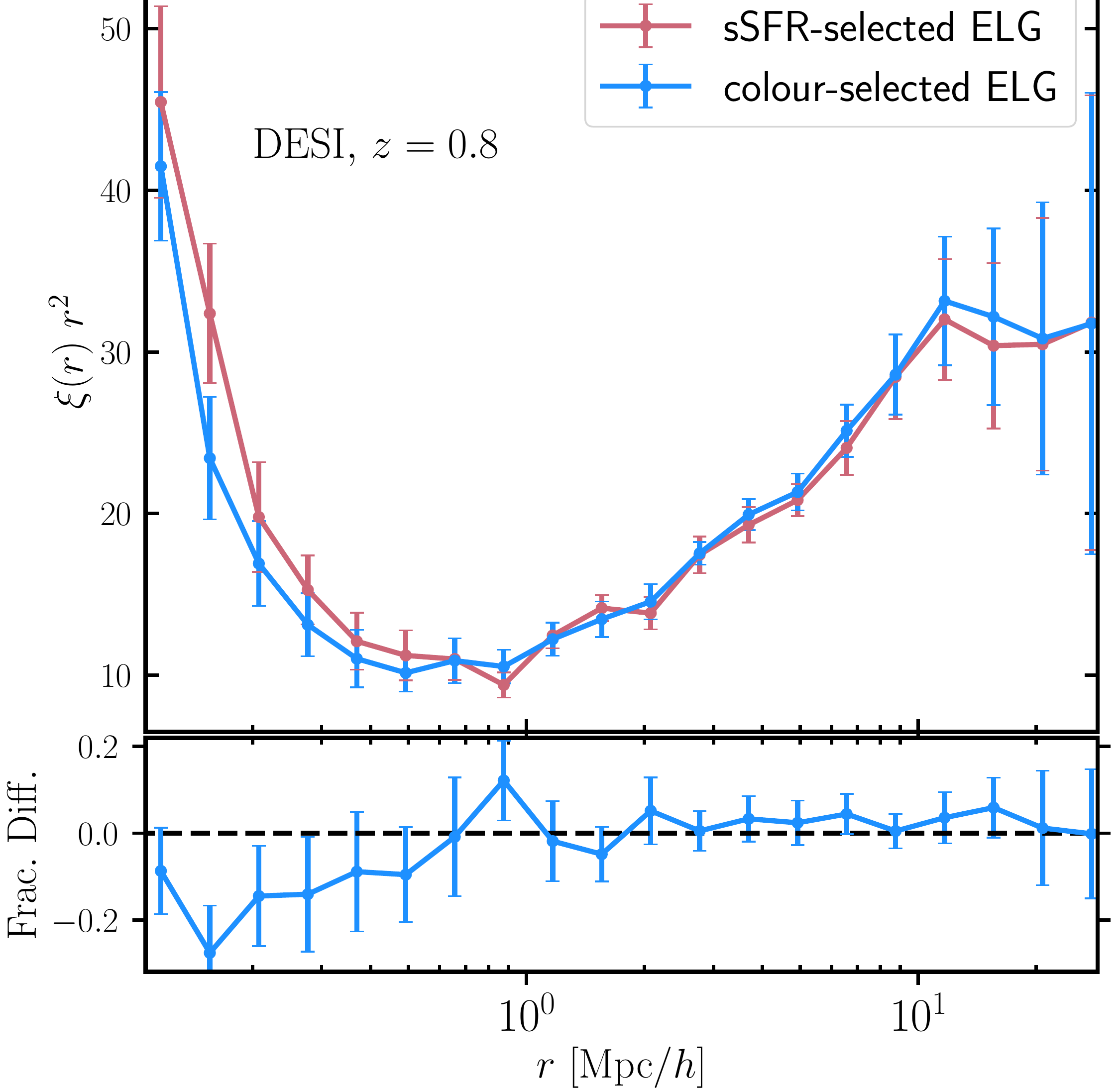}
\includegraphics[width=.48\textwidth]{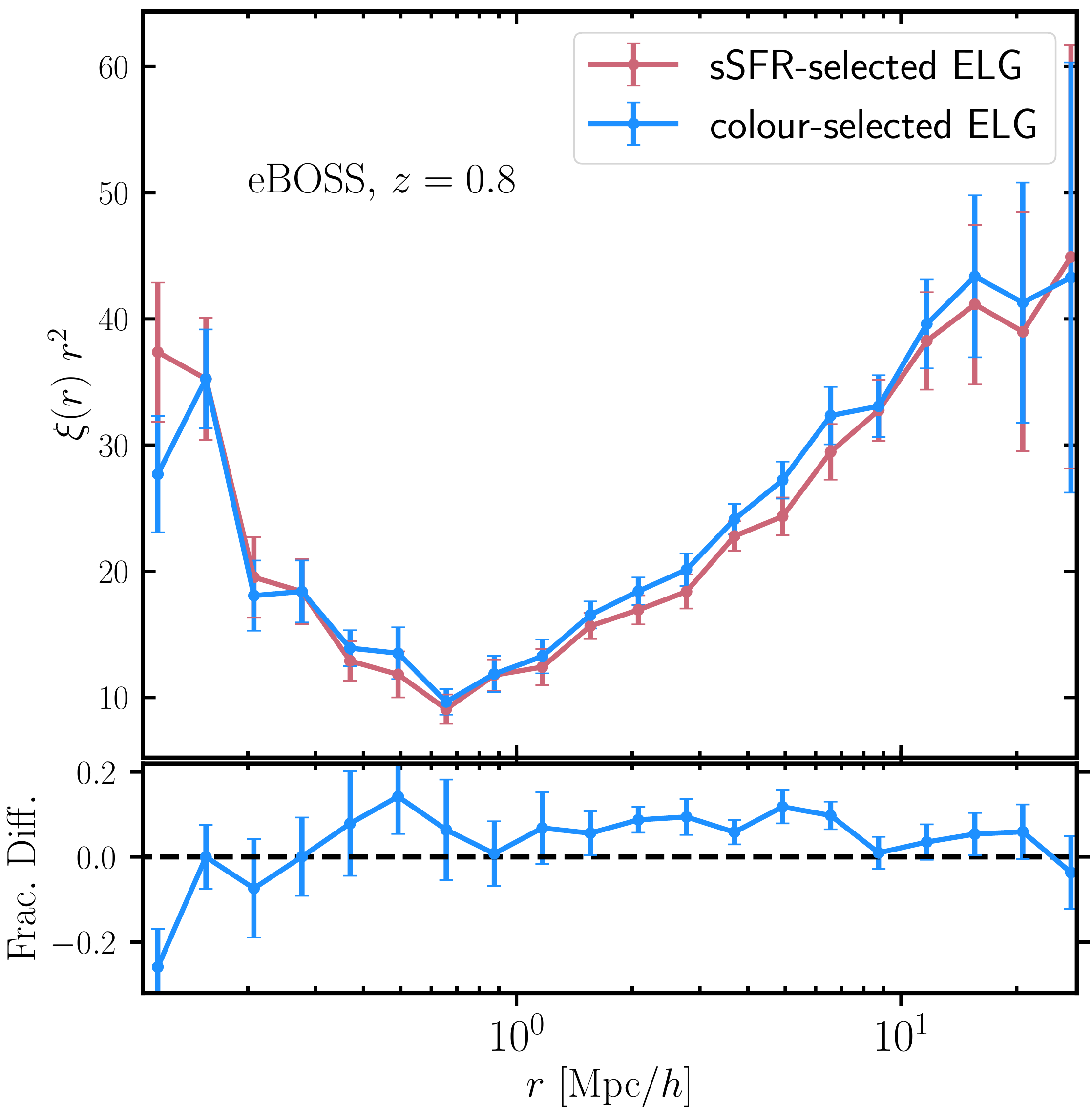} \\
\includegraphics[width=.48\textwidth]{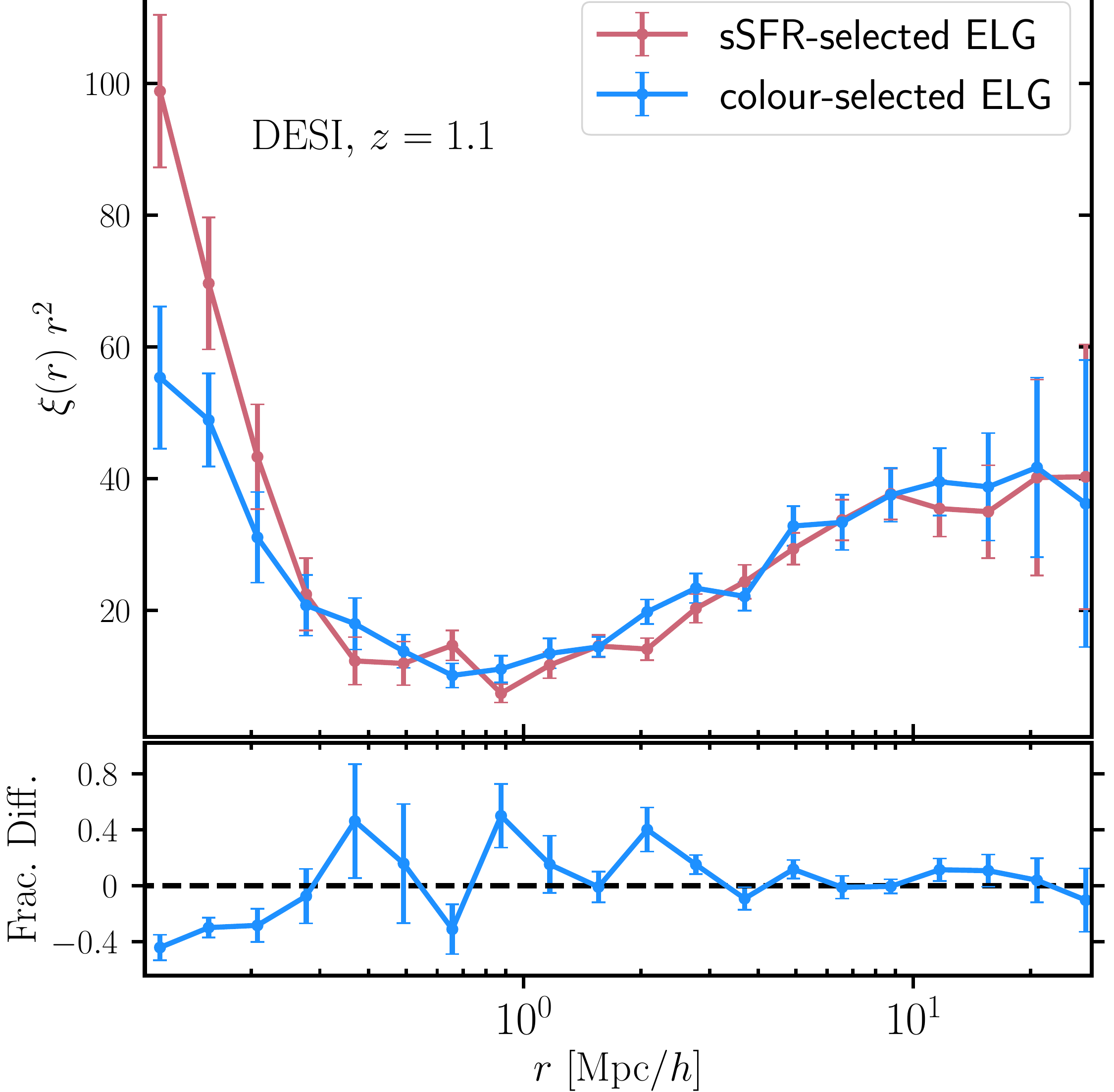}
\includegraphics[width=.48\textwidth]{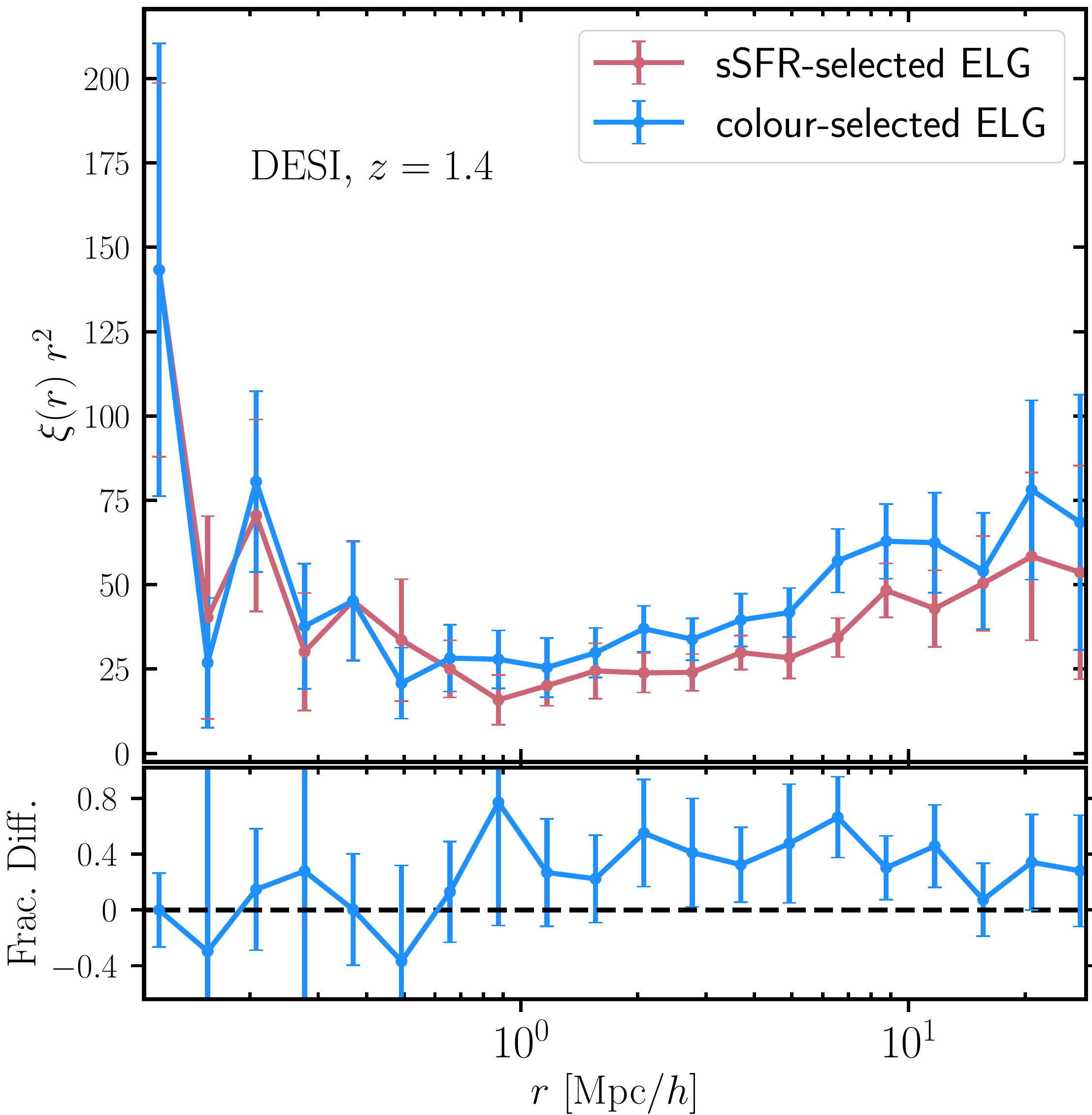}
\caption{Clustering of the
colour- (in \textit{blue}) and sSFR- (in \textit{red}) selected ELGs. The colour selections are done using the DESI (\textit{top left} and \textit{bottom panels}) and eBOSS proposed cuts (\textit{top right panel}), detailed in Table \ref{tab:obs}, while the procedure for obtaining the sSFR sample is laid out in Section \ref{sec:meth.sfr} {and Table~\ref{tab:sfr}}. In each of the four plots, the upper section shows the two-point correlation function, $\xi(r) r^2$, while the lower section shows the fractional difference in the clustering relative to the sSFR-selected objects. The eBOSS sample presented uses the $z = 0.8$ TNG galaxies, while the samples for DESI are taken at $z = 0.8$, $z = 1.1$, and $z = 1.4$. We see that the agreement between the colour- and sSFR-selected samples is best for the DESI $z = 0.8$ sample and worsens with redshift. The largest discrepancies are seen on small scales ($\sim~1 \ {\rm Mpc}/h$). Significant differences on large scales of about 50\% can be observed at $z = 1.4$, where the colour-colour distribution of galaxies has shifted towards the DESI ELG-targeting boundaries and a larger number of red quiescent galaxies are selected (see Fig.~\ref{fig:deep2} and Fig.~\ref{fig:hod}).}
\label{fig:corr}
\end{figure*}

The spatial two-point correlation function, $\xi(r)$, measures
the excess probability of finding a pair of galaxies at a given
separation, $r$, with respect to a random distribution. It is a central
tool in cosmology for studying the three-dimensional distribution
of objects and thus constraining cosmological parameters. We compute the 
two-point correlation function of the galaxy samples using the
{natural estimator \citep{1980lssu.book.....P}}:
\begin{equation}
    \xi_{\rm LS}(r) = \frac{DD(r)}{RR(r)}-1 
\end{equation} 
via the package \textsc{Corrfunc} \citep{2020MNRAS.491.3022S}
assuming periodic boundary conditions.
We estimate the uncertainties of the correlation function using
jackknife resampling \citep{2009MNRAS.396...19N}, 
dividing the simulation volume into 27 equally sized boxes
and adopting the standard equations:
\begin{equation}
    {\bar \xi(r)}=\frac{1}{n}\sum_{i=1}^{n} {\xi}_i(r)
    \end{equation}
    \begin{equation}
    {\rm Var}[{\xi(r)}]=\frac{n-1}{n} \sum_{i=1}^{n} ({\xi_i(r)} - {\bar \xi(r)})^2 ,
\end{equation}
to calculate the mean and jackknife errors of the correlation 
functions, where $n=27$ and ${\xi_i(r)}$ is the correlation
function value at distance $r$ for subsample $i$ (i.e. excluding the
galaxies residing within volume element $i$ in the correlation function computation).

In Fig.~\ref{fig:corr}, we show the clustering of the ELG-like samples using the
various selection criteria outlined in Section \ref{sec:meth}. In \textit{blue}, we show the
colour-selected samples using the DESI (top left and bottom panels) and eBOSS
(top right) cuts (see Table \ref{tab:obs}), while in \textit{red} we show the
sSFR-selected samples at the same galaxy number density (see Section \ref{sec:meth.sfr} {and Table~\ref{tab:sfr}}
for a description of the sSFR selection and Table \ref{tab:num} stating the number densities). The top panels show the redshift samples
at $z = 0.8$ and the bottom panels correspond to $z = 1.1$ and $z = 1.4$ (left and
right, respectively). Each panel displays the two-point correlation function $\xi(r) r^2$
and beneath it the ratio of the colour-selected sample
with respect to the sSFR-selected one.

We see that the agreement between the two ELG-selection strategies
(i.e. based on colour cuts and sSFR cuts) is very good for all cases, although the
deviations are more apparent for the higher redshift samples. For the DESI $z = 0.8$
panel, the largest discrepancies appear on small scales, but the
difference is consistent with zero on large scales. The top right panel demonstrates 
that the eBOSS ELGs are biased high compared with the sSFR-selected sample. The behavior in the bottom right panel, which displays the $z = 1.4$ sample, is similar.
These findings can be explained by considering the right panels of Fig.~\ref{fig:hod},
which show that the colour-selected samples find slightly more centrals than the
sSFR-selected ones (and also fewer satellites since the number density is constant between
the two). Fig.~\ref{fig:deep2} hints at an explanation of this effect: with increasing redshift the galaxies colour-colour distribution moves towards the ELG target selection boundaries, allowing a larger number of red quiescent galaxies to enter the sample. As a result, the clustering on large scales increases since the presence of a larger number
of centrals contributes to the two-halo term. Conversely, a larger number of satellites
enhances the one-halo term.

\subsubsection{Environment dependence of ELG clustering}
\begin{figure}
\centering  
\includegraphics[width=1.\columnwidth]{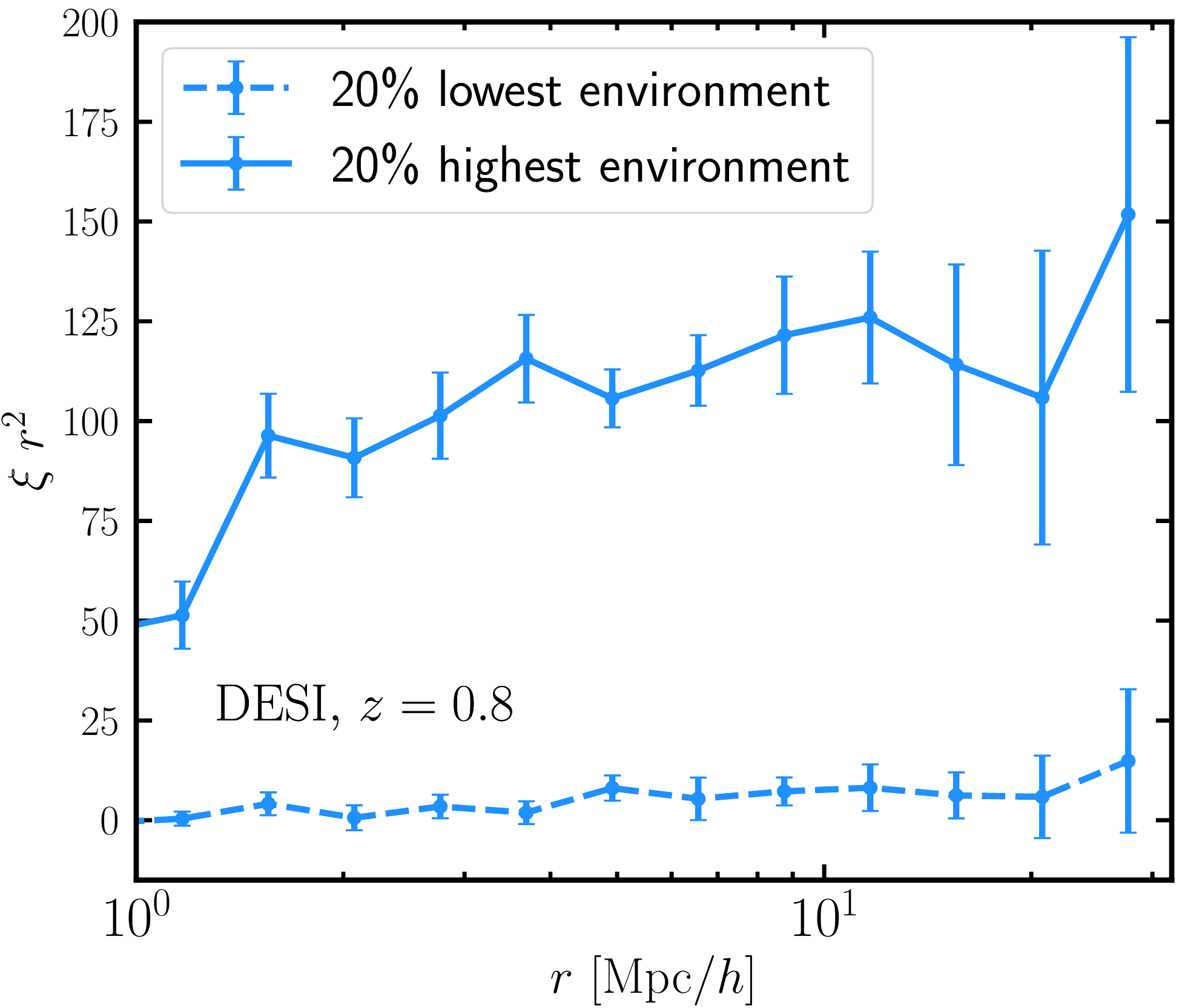} \\
\includegraphics[width=1.\columnwidth]{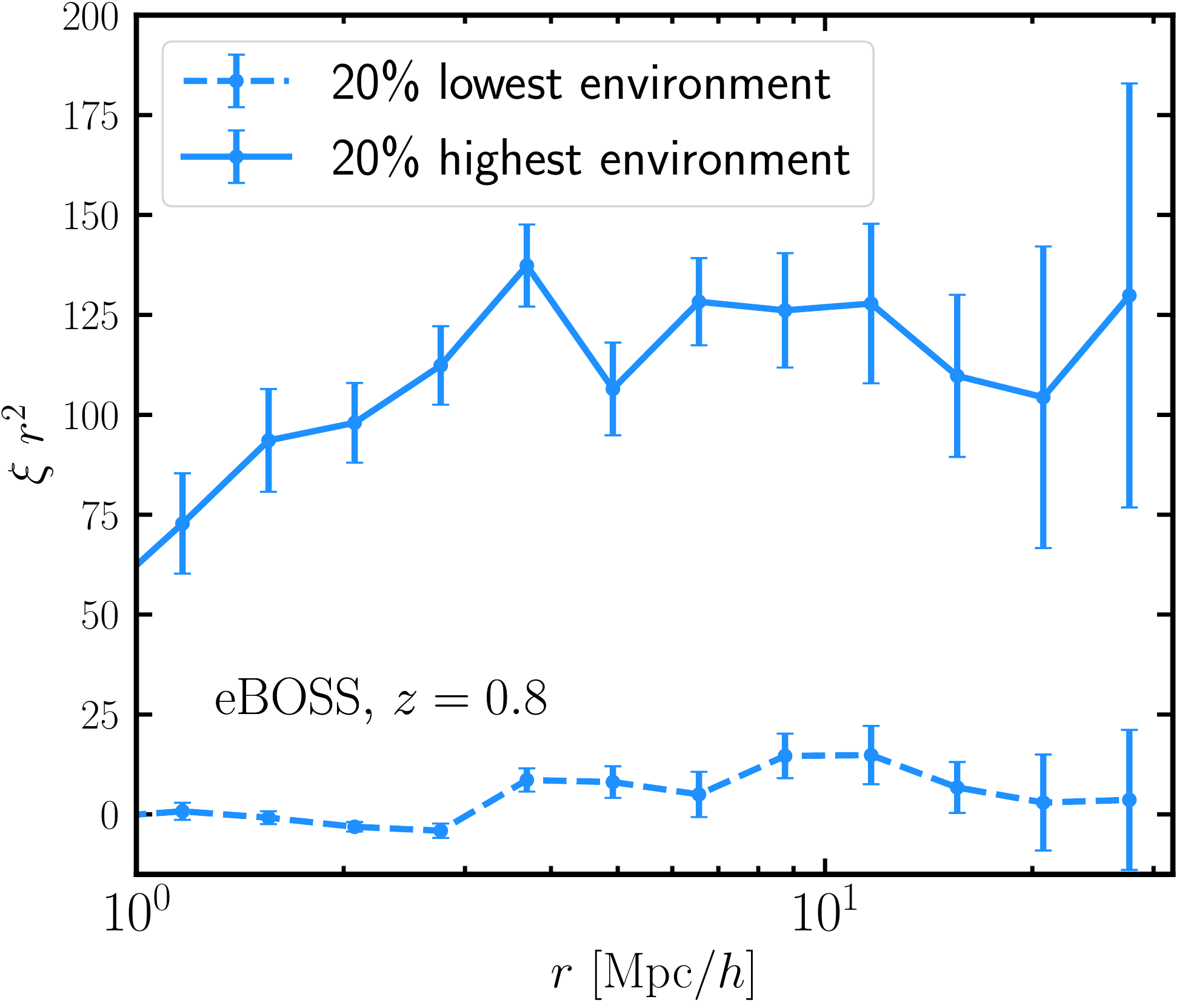}
\caption{Clustering of the colour-selected ELG-like objects in high (\textit{solid lines}) and low (\textit{dashed lines}) environments at redshift $z = 0.8$. The colour-selected sample in the \textit{top panel} uses the DESI proposed cuts, while the one in the \textit{bottom panel} uses the eBOSS cuts (see Table \ref{tab:obs}). Each plot shows the two-point correlation function, $\xi(r) r^2$ for both environments. For each halo mass bin, we haven take the top and bottom 20\% haloes ordered by their environment factor (as defined in Section \ref{sec:res.hod.env}) and computed the correlation functions of the galaxies living in each of the two subsample. We see that the galaxies living in the higher density environments exhibit markedly stronger clustering than their lower density counterparts, which is important to model corectly. Here we define halo environment as the Gaussian-smoothed matter density over a scale of $R_{\rm smooth} = 1.4 \ {\rm Mpc}/h$.}
\label{fig:corr_env}
\end{figure}

In Fig.~\ref{fig:corr_env}, we show the dependence of the galaxy clustering on
environment. We split the galaxies into belonging to high- and low-density
environments using a similar approach to that adopted in Section \ref{sec:res.hod.env}
for obtaining an HOD augmented with an environment parameter. For each halo mass
bin of size $\Delta \log(M_{\rm halo}) = 0.1 \ {\rm dex}$, we have chosen the top and bottom 
20\% haloes ordered by their environment factor (as defined in Section 
\ref{sec:res.hod.env}). We then compute the two-point clustering of the galaxies living inside them. We display these results in Fig.~\ref{fig:corr_env} for the
$z = 0.8$ redshift sample, as it hosts a larger number of objects 
(see Table \ref{tab:num}). As before, we define halo environment as
the Gaussian-smoothed matter density over a scale of $R_{\rm smooth}
= 1.4 \ {\rm Mpc}/h$. 

The two panels show the clustering, $\xi(r) r^2$, using the DESI and
eBOSS selection criteria, respectively (see Table \ref{tab:obs}).
The clustering of the low-environment galaxies is denoted by
a \textit{dotted} line, while that of the high-environment ones by a \textit{solid} line.
We see that the galaxies living
in denser environments are more strongly clustered than their low-density counterparts,
suggesting that at fixed mass, clustering is dependent on additional
assembly bias properties pertaining to the halo 
environment. Indeed, recognition of the importance of this
effect has motivated a number of recent ``augmented'' HOD 
models that, in addition to halo mass, also incorporate 
secondary and tertiary parameters in the galaxy-halo 
connection, such as environment and concentration 
assembly bias \citep[e.g.][]{2020MNRAS.493.5506H,2020Xu,2020arXiv201004182Y}.

\subsection{Bias and correlation coefficient}
\label{sec:res.bias}
Most of the cosmological information of the matter 
distribution is encoded in the power spectrum (or correlation function)
of the matter density fluctuations as a function
of scale and redshift. However, galaxies are not perfect tracers of the
underlying mass distribution, and thus, it is 
important to understand the relationship between the 
large-scale distribution of matter and that of galaxies. 
The galaxy auto-correlation function, $\xi_{gg}(r)$ is related to the
matter correlation function, $\xi_{mm}(r)$, through
the real-space galaxy bias, $\tilde b$,
in the following way:
\begin{equation}
    \xi_{gg}(r) = \tilde b^2(r) \xi_{mm}(r).
\end{equation}
One can furthermore study the bias through the
galaxy-matter cross-correlation function,
$\xi_{gm}(r)$, which can be related to the matter 
two-point correlation function through $\tilde b$
and the real-space cross-correlation 
coefficient between matter and galaxy fluctuations, 
$\tilde r$ \citep{2008MNRAS.388....2H,2018PhR...733....1D}:
\begin{equation}
    \xi_{gm}(r) = \tilde b(r) \tilde r(r) \xi_{mm}(r)
\end{equation}
where the galaxy bias is:
\begin{equation}
    \tilde b(r) = \Bigg[\frac{\xi_{gg}(r)}{\xi_{mm}(r)}\Bigg]^{\frac{1}{2}}
\end{equation} 
and the correlation coefficient is:
\begin{equation}
    \tilde r(r) = \frac{\xi_{gm}(r)}{[\xi_{gg}(r) \ \xi_{mm}(r)]^{1/2}} .
\end{equation} 
The equations above are general and may be taken as
definitions of the scale-dependent galaxy bias 
$\tilde b(r)$ and cross-correlation coefficient 
$\tilde r(r)$. We note that the quantity $\tilde 
r(r)$ in real space is not constrained to be less 
than or equal to one. However, one expects $\tilde r(r)$
to approach unity  on  large  scales,  where  the 
observed correlation should be sourced from the gravity field of the total matter. 

\begin{figure}
\centering  
\includegraphics[width=.5\textwidth]{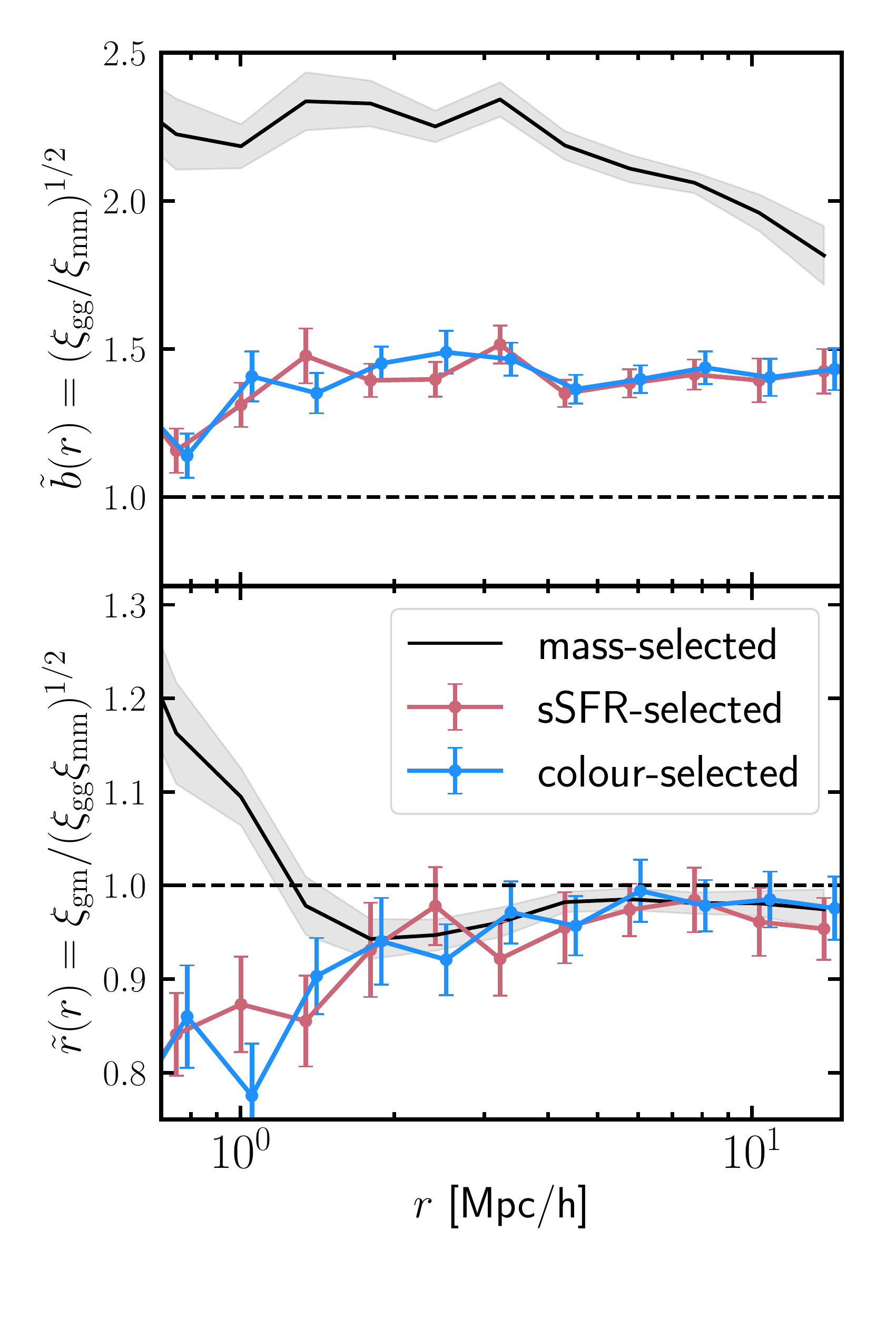}
\caption{Bias and correlation coefficient at $z = 0.8$ for the following three samples: one selected based on a stellar-mass
cut (shown in \textit{grey solid}), one selected using the proposed
DESI cuts (shown in \textit{blue solid} and described in Section \ref{sec:meth.desi}), and one selected using the 
sSFR cuts (shown in \textit{red solid} 
and described in Section \ref{sec:meth.sfr} {and Table~\ref{tab:sfr}}). \textit{Top panel}
shows the galaxy bias, $\tilde b(r)$, defined as the 
square root of the ratio of the galaxy and
matter auto-correlation functions, while the 
\textit{bottom panel} shows the real-space 
correlation coefficient, $\tilde r(r)$. The
galaxy bias goes to a constant on large scales,
and the correlation coefficient approaches 1,
suggesting that a linear bias approximation on
scales beyond 10 Mpc$/h$ is appropriate. We see
that the agreement between the galaxy bias 
of the sSFR-selected sample and that of the ELG sample
is excellent, as expected from the two-point clustering plots,
and their real-space cross-correlation coefficients
are also compatible with each other.
On the other hand, the mass-selected sample has a
substantially higher galaxy bias ($\tilde b(r) \approx 2.1$)
than the ELG-like samples ($\tilde b(r) \approx 1.4$),
and the cross-correlation coefficient between galaxies
and matter is also slightly higher.}
\label{fig:bias_corr}
\end{figure}

In Fig.~\ref{fig:bias_corr}, we demonstrate what these look
like for the colour-selected sample using the DESI ELG targeting 
criteria (see Table \ref{tab:obs}), the sSFR-selected sample (see Section
\ref{sec:meth.sfr} {and Table~\ref{tab:sfr}} for a detailed description of the selection procedure),
and a mass-selected sample matching the galaxy number density of the other two 
(stated in Table \ref{tab:num}).
The TNG redshift sample used in this figure corresponds to $z = 0.8$. 
From the top plot, we can infer that the galaxy bias roughly approaches 
a constant on large scales for the sSFR- and colour-selected samples,
where $\tilde b(r) \approx 1.4$, while the bias of the mass-selected sample
is much higher, $\tilde b(r) \approx 2.1$. This difference in the bias of the
two samples is most likely a consequence of the fact that the most massive galaxies
live predominantly in the densest regions (knots, see Fig.~\ref{fig:2d_distn}), 
while the ELG-like galaxies are more likely to be found in filamentary regions,
which are less strongly biased \citep[see e.g.][for a review on Excursion Set Theory]{2007Zentner}. We find the average bias on large scales ($1 \ {\rm Mpc}/h < r < 20 \ {\rm Mpc}/h$) for the other two redshift samples to be $\tilde b(r) \approx 2.3, \ 1.5, \ {\rm and} \ 1.5$ for the mass-, sSFR- and colour-selected samples, respectively, at $z = 1.1$, and $\tilde b(r) \approx 2.9, \ 1.7, \ {\rm and} \ 2.0$ for the mass-, sSFR- and colour-selected samples, respectively, at $z = 1.4$.
The fact that on large scales the galaxy bias tends to a constant value suggests
we can use the linear bias approximation to infer the underlying matter distribution
\citep{1980lssu.book.....P,1996MNRAS.282..347M,2013MNRAS.432.1544M}.
For the linear bias approximation to be valid,
the cross-correlation coefficient needs also be scale-independent on large scales, 
approaching unity \citep{2010PhRvD..81f3531B,2020PhRvD.102h3520S}. We can see from the lower panels
that this is indeed the case and $\tilde r(r) \approx 1$ on large scales, implying
that as long as one considers large-scale galaxy clustering on scales
much greater than 1 Mpc, the observed correlation
should be sourced from the gravity field of the total matter. 

However, it is important to note that while on large scales
the linear bias approximation appears to be viable, it certainly
breaks down on smaller scales ($\sim$1 Mpc$/h$). This 
has important implications for analyses using mock catalogues created
via phenomenological approaches such as the HOD framework. The small-scale
signal encodes a lot of information about cosmological parameters such as
$\Omega_m$ and $\sigma_8$. In addition, modeling these scales correctly is a key
requirement for weak lensing shear analysis. Finally, the small-scale
data provide an important window for probing different DM models and 
understanding the effects of baryonic physics.
It is reassuring to see that the galaxy bias is in a 
very good agreement between the two ELG-like samples (\textit{red} and \textit{blue})
in Fig.~\ref{fig:bias_corr}. Furthermore, the cross-correlation coefficients 
derived for the ELG-like samples exhibit a very similar behavior across all scales,
suggesting that the galaxy-matter cross correlation relates
similarly to the galaxy and matter clustering regardless
of the underlying population model. The minimum point of
the cross-correlation coefficient is around ($r\sim~1$ Mpc$/h$), which
corresponds to the outskirts of haloes, between the one- and two-halo 
terms, where the dark matter outweighs the luminous component, which has sunk to
the halo center, as it radiates gas and dissipates energy. Note that the mass-selected galaxies
tend to be larger and therefore the border between the one- and two-halo terms
is pushed back to slightly larger scales ($r\sim~2$ Mpc$/h$).

\section{Mock catalogues}
\label{sec:mock}
In this section, we provide motivation for an HOD 
population mechanism that incorporates environment
by measuring the galaxy assembly bias of the DESI ELG-like
sample as well as the sSFR- and mass-selected ones. 
We then analyse the satellite distributions of the DESI
ELGs and compare it with that of the subhaloes rank-ordered by
6 different subhalo properties. Finally, we propose a
detailed procedure for obtaining a mock catalogue incorporating
an environmental dependence and show its success in recovering
the large-scale clustering of the various samples examined
in this work.

\subsection{Galaxy assembly bias}
\label{sec:mock.gab}

A standard way for assessing the amount of assembly bias
of a given galaxy sample is to compare the two-point correlation
function of the sample with that of a shuffled sample, where
the galaxy occupation numbers of haloes are randomly reassigned
to haloes belonging to the same mass bin \citep{2007MNRAS.374.1303C}. 
The goal of this shuffling procedure is to erase the link between
the halo occupation and assembly history, thus eliminating the
dependence on any secondary properties other than halo mass.
Since in this section we are only interested in measuring
the large-scale galaxy assembly bias, we do not need to adopt
a realistic prescription for populating haloes on small (intra-halo) scales.
Instead, when we transfer galaxies from their original halo
into their newly assigned host, we move their locations
in a way which maintains their relative positions with respect
to the halo centre (defined as the spatial position of the particle
with the minimum gravitational potential energy). 
When shuffled in this way, the contribution of the one-halo
term to the auto-correlation function is preserved. 
The size of the halo mass bins within which we shuffle
the halo occupations is $\Delta \log(M_{\rm halo}) = 0.1 \ {\rm dex}$. 

\begin{figure}
\centering  
\includegraphics[width=.5\textwidth]{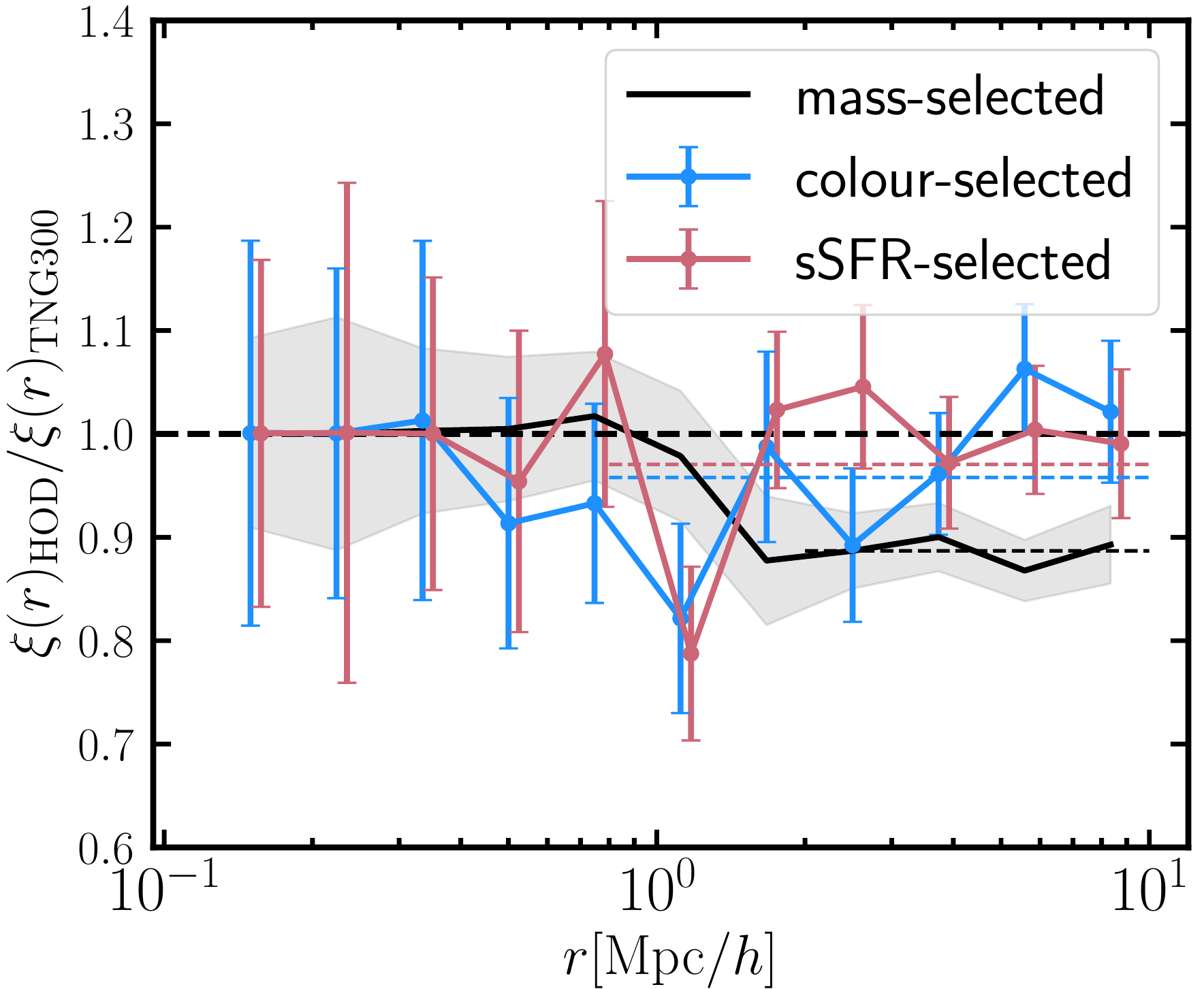}
\caption{A demonstration of the assembly bias effects when comparing
the ratio between shuffled samples (with the relative galaxy positions
preserved) as outlined in Section \ref{sec:mock.gab} and
the ``true'' galaxy sample for the following three shuffled
samples: one selected based on a stellar-mass
cut (shown in \textit{black solid}), one selected using the proposed
DESI cuts (shown in \textit{blue solid}), and one selected using the 
sSFR cuts (shown in \textit{red solid}). In \textit{dotted lines},
we show the large-scale values for the three cases
averaged between $r = 0.8 \ {\rm Mpc}/h$ ($r = 2 \ {\rm Mpc}/h$ for the mass-selected sample) and 10 Mpc$/h$. We see that
the mass-selected sample exhibits the largest amount of discrepancy,
$\sim$10-15\% in agreement with previous works \citep[e.g.][]{2020MNRAS.493.5506H,2019arXiv190811448B},
while the colour- and sSFR-selected samples show much smaller deviations.
This implies that the effect of assembly bias is likely smaller for
ELG-like objects.}
\label{fig:shuff}
\end{figure}

In Fig.~\ref{fig:shuff}, we present the results of 
performing the shuffling procedure applied to three galaxy
samples of interest (preserving the relative positions of the galaxies with respect to the halo centres): a DESI-selected ELG sample (see Section 
\ref{sec:meth.desi}), an sSFR-selected sample (see Section 
\ref{sec:meth.sfr} {and Table~\ref{tab:sfr}}), and a mass-selected sample, where we
apply a stellar mass cut to the galaxies chosen in a way
so that the number of galaxies in all three cases is equal.
Here we apply this analysis to $z = 0.8$, where we have the
largest expected number of ELGs (see Table \ref{tab:num}).
One can also notice that the clustering ratio of the
mass-selected sample is consistent with 1 until slightly larger
scales than for the ELG-like samples (shown in \textit{blue} and
\textit{red}). The reason is that the mass-selected galaxies
tend to live in larger haloes, so the transition between the
one- and two-halo terms happens at larger radial distance 
($r \sim~1 \ {\rm Mpc}/h$). In dotted lines, we show the
average ratio in the range of $2-10 \ {\rm Mpc}/h$. We find that
the deviation of the stellar mass sample is the most substantial, 
at $\sim$10\%, while that of the other two samples is within
$5\%$. This suggests that while assembly bias imparts a smaller 
effect on ELGs, careful modeling is still necessary to achieve the 
required amounts of precision. 

\subsection{Satellite distribution}
\label{sec:mock.small}

\begin{figure*}
\centering
\includegraphics[width=.48\textwidth]{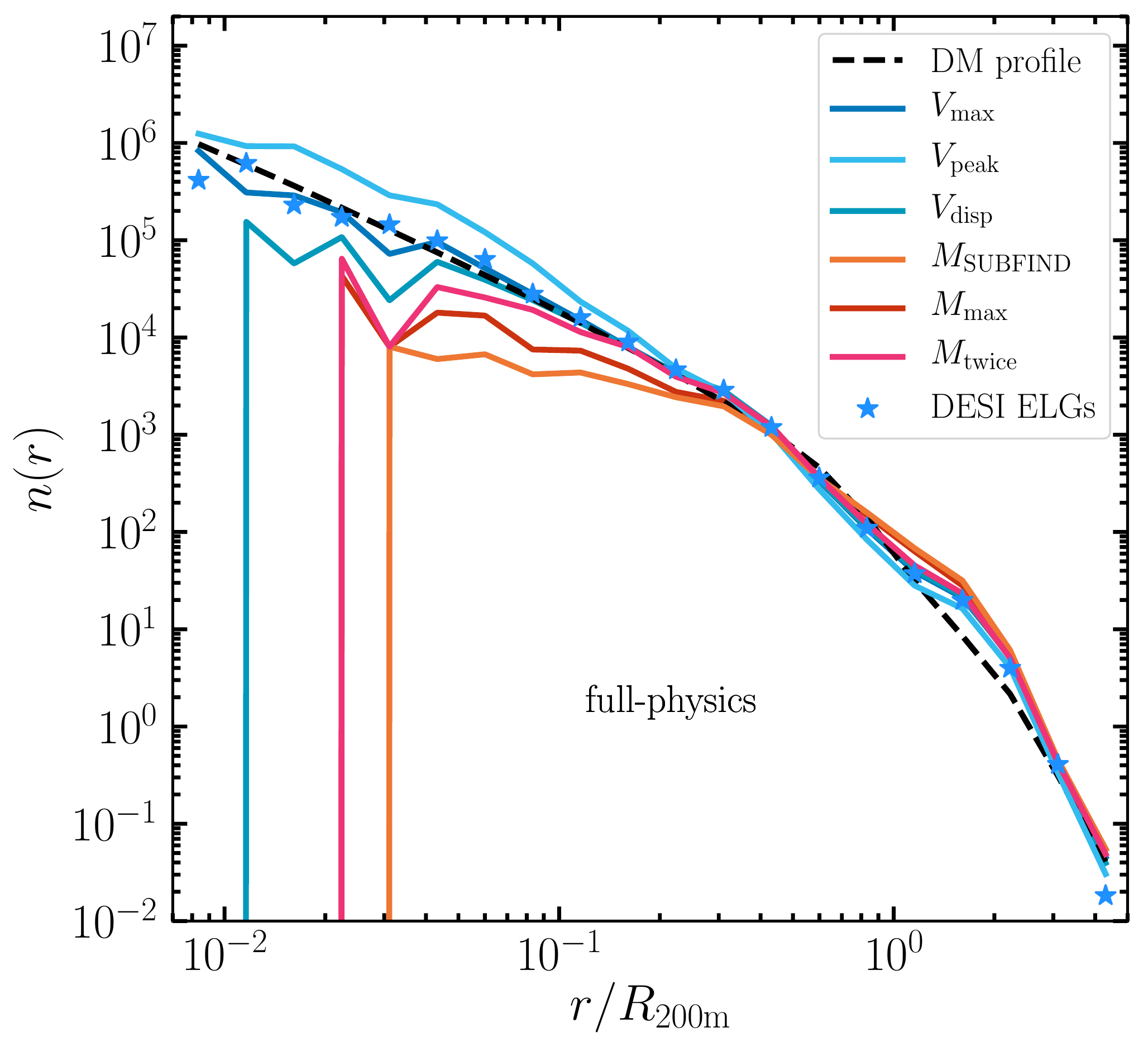}
\includegraphics[width=.48\textwidth]{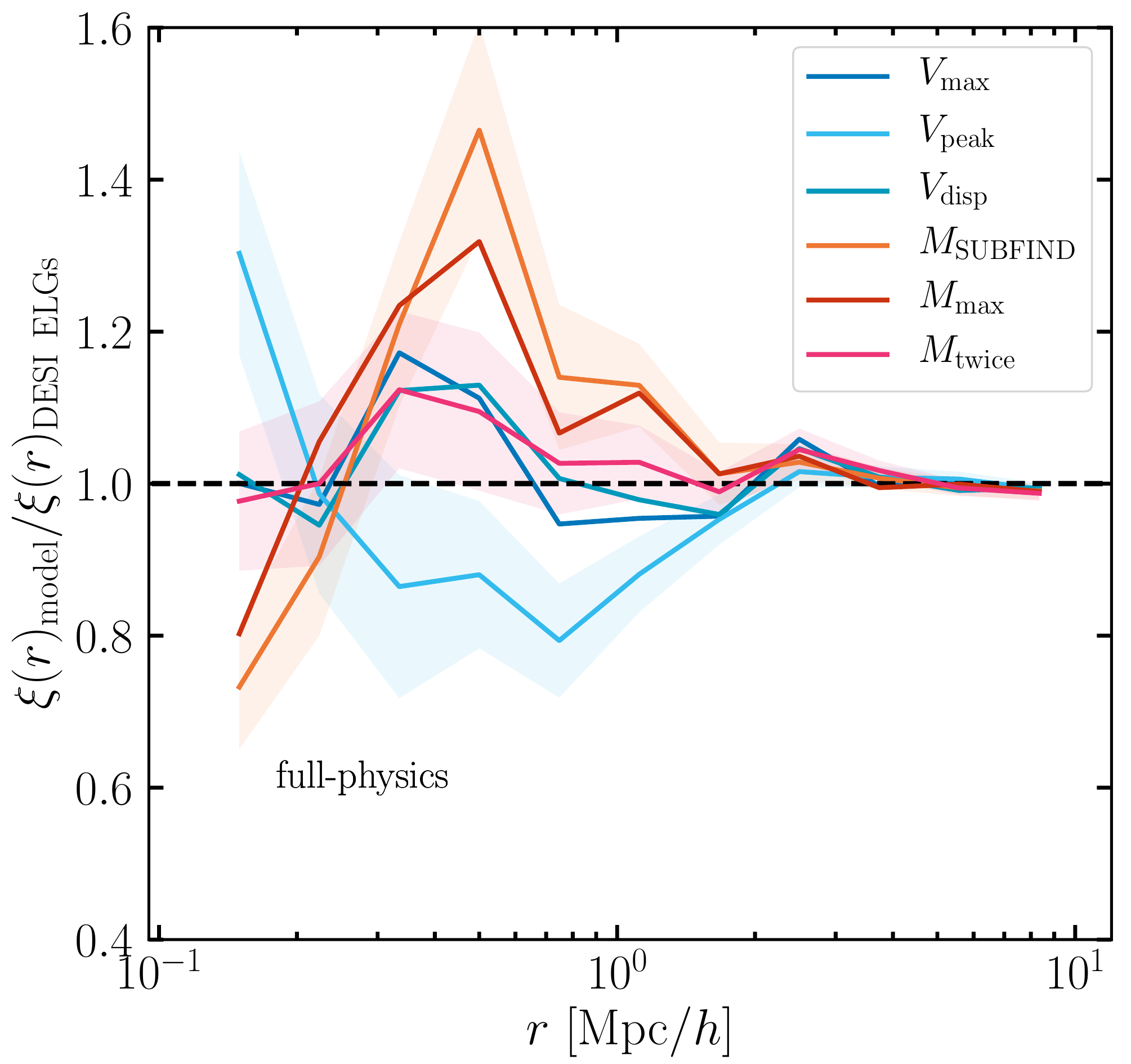} \\
\includegraphics[width=.48\textwidth]{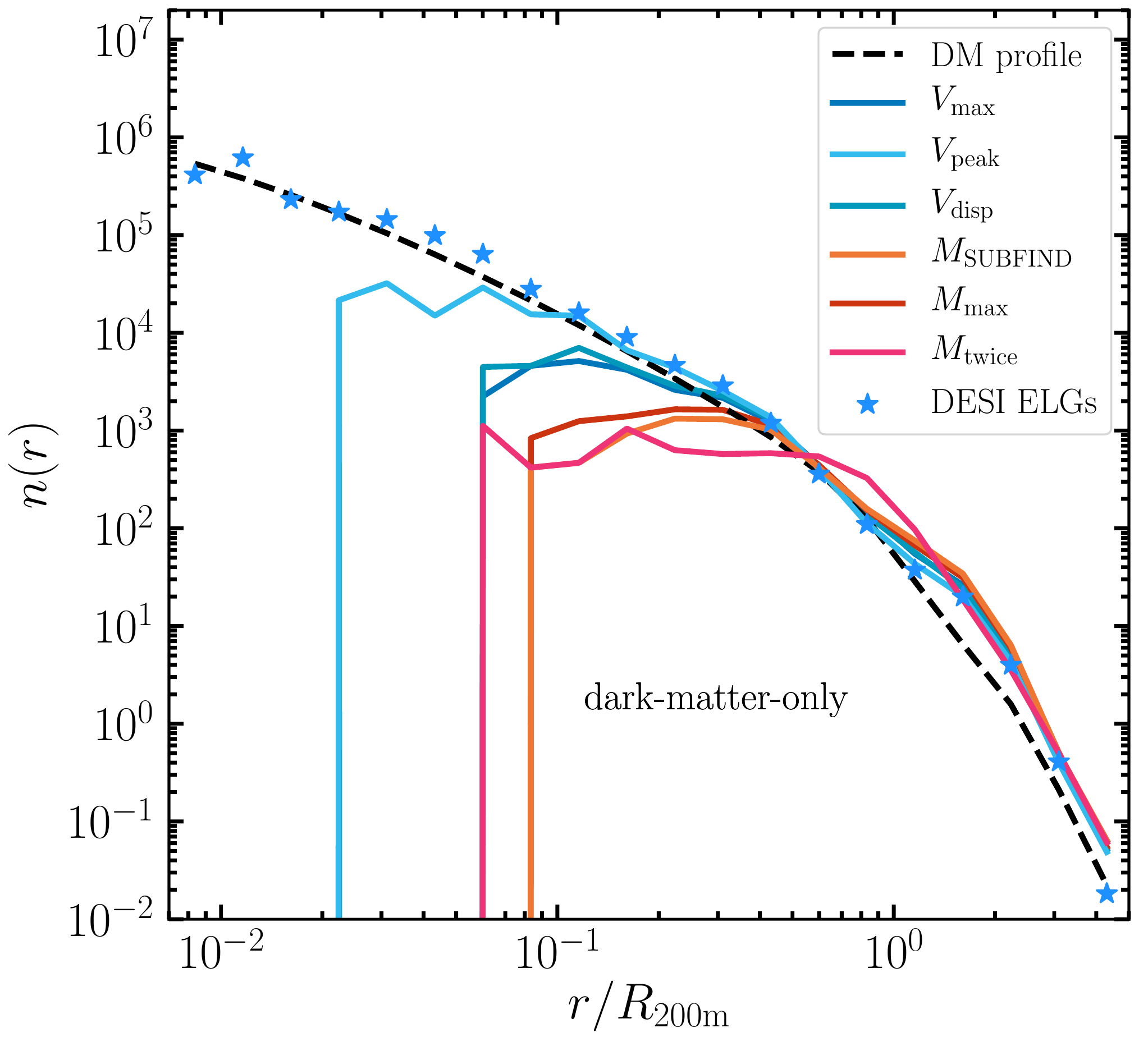}
\includegraphics[width=.48\textwidth]{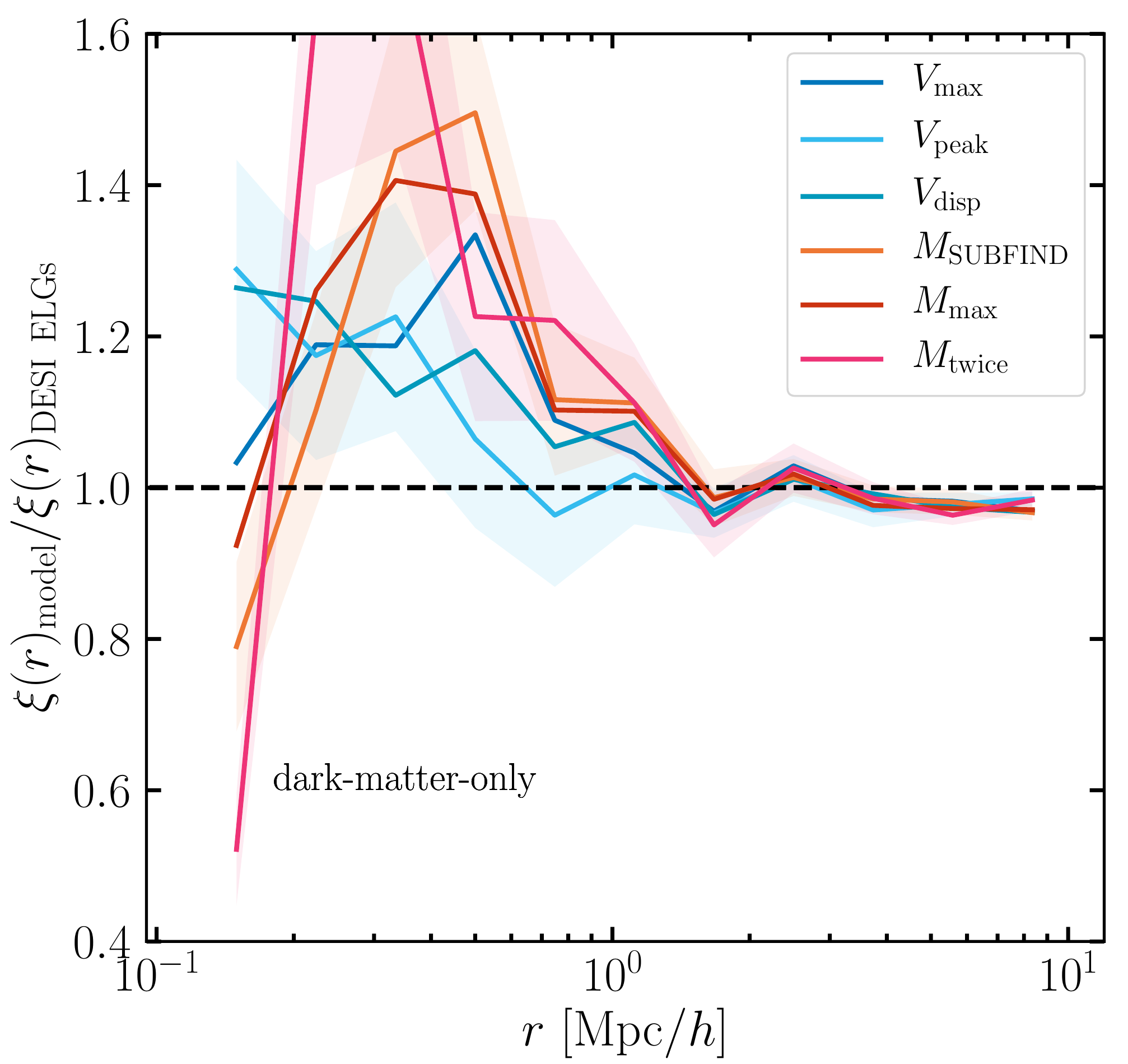}
\caption{\textit{Left panels}: Number density profile as a function of distance from the halo centre for the DESI colour-selected ELGs at $z = 0.8$ (see Table \ref{tab:obs}) and the subhaloes selected based on several velocity- and mass-based properties: $V_{\rm max}$, $V_{\rm peak}$, $V_{\rm disp}$, $M_{\rm SUBFIND}$, $M_{\rm max}$, and $M_{\rm twice}$ (see definitions in Section \ref{sec:mock.small}). The radial distance is measured in units of $R_{\rm 200m}$, the halo radius at 200 times the mean density. For reference, we also show the number density profiles of the dark matter particles in \textit{dashed black}. \textit{Right panels}: Clustering ratio of samples populated by rank-ordering the six subhalo properties with respect to the ``true'' ELG positions. The shaded regions denote the $1\sigma$ errors. The \textit{top two panels} refer to the full-physics run of TNG300 (FP), while the \textit{bottom panels} refer to the dark-matter-only (DMO) run of the same box. In the \textit{top left} panel, we see that $V_{\rm max}$ matches the satellite distribution of ELGs best over all scales. The other 5 parameters diverge on smaller scales ($r/R_{\rm 200m} \lesssim 0.3$) with $V_{\rm peak}$ and $V_{\rm disp}$ being the next best-performing parameters. However, these small-scale differences have little effect on the clustering, which we study on scales of $10^{-1} {\rm Mpc}/h < r < 10 {\rm Mpc}/h$. We also see that the dark matter profile (\textit{dashed black}) follows the ELG distribution quite well. The \textit{top right panel} demonstrates that the full-physics ELG clustering is matched better (within $\sim$15\%) by $V_{\rm max}$, $V_{\rm disp}$, and $M_{\rm twice}$, compared with the other three parameters. On large scales, the clustering of all samples agrees with that of the ELGs. This is as expected since the choice of a mechanism for populating the one-halo term should not affect the two-halo term clustering. In the \textit{bottom left panel}, we have illustrated that for all parameters, the satellite distributions in the dark-matter-only simulation tend to flatten on small scales ($r/R_{\rm 200m} \lesssim 10^{-1}$) and are unable to provide a fit to the number density profile of DESI ELGs in that regime. This is least the case for the $V_{\rm peak}$ parameter followed by $V_{\rm disp}$ and $V_{\rm max}$, which appear to be steeper than the mass-based parameters on these scales. On the scale of interest for the clustering comparison ($r > 10^{-1} {\rm Mpc}/h$), the parameters that exhibit the most evident deviation are the mass-based ones ($M_{\rm SUBFIND}$, $M_{\rm max}$, $M_{\rm twice}$), while the other three behave similarly. Similarly to the full-physics case, the dark matter profile (\textit{dashed black}) is in close agreement with the ELG distribution. In the \textit{bottom right panel}, we see that the dark-matter-only subhalo parameter displaying the smallest discrepancy from the ELG clustering is $V_{\rm peak}$ followed by $V_{\rm disp}$ and $V_{\rm max}$. It is interesting to notice that in all cases, the small-scale clustering is slightly overpredicted when placing the galaxies in a halo according to a rank-ordered list.}
\label{fig:prof}
\end{figure*}

One of the key components of modeling the galaxy-halo connection
is deciding upon an algorithm to internally assign galaxies
in haloes, i.e. once the number of centrals and satellites has
been picked, one needs to determine how the satellites are distributed
spatially within their host haloes. This is vital for recovering
accurate small-scale clustering on scales of $\lesssim 1 \ {\rm Mpc}/h$
(the one-halo term). 

Typically, satellite positions 
are assigned by one of three commonly-adopted schemes: (i) assuming that satellites trace the dark-matter
profile of the halo, to which one typically fits an NFW curve
and mimics its shape through the satellites; (ii) placing satellites
on a randomly-selected dark matter particle;
(iii) assuming they follow the radial distribution of the
dark-matter-only subhaloes (typically adopting an abundance matching
technique conditioned on some subhalo property). 
In hydrodynamical simulations, the knowledge of the ``true'' positions
of the galaxies is provided since the simulations are evolved
self-consistently, accounting for baryonic physics. We can
therefore compare each of these methods before deciding
upon the best strategy to populate the one-halo term. For the abundance matching step, we select the locations of satellite galaxies according to the following set of subhalo properties:
\begin{itemize}
    \item $V_{\rm max}$, the maximum circular velocity of the subhalo at the final time ($z = 0.8$),
    \item $V_{\rm peak}$, the maximum circular velocity the subhalo reaches throughout its history,
    \item $V_{\rm disp}$, the dispersion velocity of the subhalo,
    \item $M_{\rm SUBFIND}$, the total mass of all bound particles in the subhalo as identified by SUBFIND \citep{Springel:2000qu},
    \item $M_{\rm max}$, the total mass (of all components) of the subhalo contained within the radius at which it attains its maximum circular velocity,
    \item $M_{\rm twice}$, the total mass (of all components) of the subhalo contained within twice the stellar halfmass radius.
\end{itemize}

In the two \textit{left panels} of Fig.~\ref{fig:prof}, we present the 
radial profiles of the satellite ELGs found by applying the DESI
cuts to the $z = 0.8$ snapshot of the TNG300. In particular, 
we show the radial number density of ELG satellites and compare
it with the radial number density obtained by abundance matching
the subhaloes in the full-physics (\textit{top left}) and dark-matter-only
(\textit{bottom left}) simulations. For reference, we also include a curve (in dashed black) that shows the density profiles of the dark matter particles residing in all haloes hosting ELG-like objects.

The abundance matching for the
subhaloes in full-physics is performed by rank-ordering
the subhaloes in each halo by one of the six properties listed
above and selecting the top $N_s$ entries, where $N_s$ corresponds
to the number of ELG satellites in the halo. For the dark-matter-only
case, we first identify the ELG hosting haloes in the full-physics
run and their counterparts in the dark-matter-only simulation.
Once we have ordered the subhaloes by a particular property, as is typically done, we add a
scatter in logarithmic space for the velocity-based properties of 
$\Delta \log(V) = 0.1$ and none for the mass-based ones. Finally, we again select the top $N_s$ subhaloes in the 
dark-matter-only halo. The radial distances in the number density
profiles are measured in units of $R_{\rm 200m}$, the halo radius
containing 200 times the mean density of the Universe. The most significant visual correspondence of the satellite distributions on the left panels with the one-halo term on the right panels can be seen around $r/R_{\rm 200m} \approx 1$, where we can identify a direct correlation between the density profiles and the clustering ratios. This observation is in agreement with the intuitive expectation that if the radial distribution of object is matched well, the one-halo term would also be consistent \citep[see][for details of the correlation function calculation]{2007ApJ...659....1Z}. 

From the left panels of Fig.~\ref{fig:prof},
we see that the ``true'' satellite distribution is best traced by the
full-physics subhaloes ordered by $V_{\rm max}$ (on all scales)
and also by the dark-matter-only subhaloes ordered by $V_{\rm peak}$ and
$V_{\rm max}$. The curves in the dark-matter-only case (\textit{bottom left panel}) are noticeably
flatter than the full-physics ones with the three velocity-based parameters appearing to visually fit the ELG radial distributions best on scales of $r/R_{\rm 200m} < 1$. An important finding is also that the dark-matter profiles visually provide a very good match to the ELG distribution. This implies that mock catalogues that ``paint'' ELGs on top of particles may provide a good approximation to the true distribution of galaxies. We further study this in the next section, where we delve into the details of creating a realistic mock catalogue.

To further test
which abundance matching property yields a one-halo term that
is consistent with the true ELG satellites, we compute the 
auto-correlation function of the ``true'' ELG sample and that of a
sample where we have preserved the occupation numbers in each
halo but have ``painted'' the galaxies on top of the
subhaloes rank-ordered by one of the six properties.
This is shown in the \textit{right panels} of Fig.~\ref{fig:prof},
which again correspond to the full-physics and dark-matter-only
runs (top and bottom panels, respectively).
The \textit{top right} panel demonstrates that the one-halo term
(i.e. $r \lesssim 1 \ {\rm Mpc}/h$)
is better matched when using one of three parameters: $V_{\rm max}$,
$V_{\rm disp}$, and $M_{\rm twice}$, compared with the rest of the parameters.
This result corroborates the findings of the
number density profiles.

In the \textit{bottom right}
panel, we can see that the parameter that performs best is $V_{\rm peak}$ followed by $V_{\rm disp}$ and $V_{\rm max}$.
It is interesting to notice that all 6 parameters slightly overpredict
the one-halo term contribution.
On large scales (i.e. $r \gtrsim 1 \ {\rm Mpc}/h$), the clustering
of all samples and in both runs agrees with that of the ELGs as expected, since 
the choice of a mechanism for populating the one-halo term should 
not affect the two-halo term clustering. There are noticeable differences
between the subhalo distributions in the full-physics and dark-matter-only
runs, which are attributable to the differences in the physical processes governing them \citep[e.g., stellar and AGN feedback, gas cooling, etc. in the full-physics TNG box; see][for a discussion]{2020arXiv200804913H}.

The analysis in this section provide a solid motivation for 
adopting a satellite population technique based on rank-ordered lists
(by a particular property) of the subhaloes or by ``painting'' galaxies onto particles in the halo, as we have seen that the dark matter profile follows closely the ELG radial distribution. In particular, when creating 
mock catalogues, (see details of our prescription in the next section),
we apply the abundance-matching model to assign the locations of satellites
within haloes based on either the subhalo property $V_{\rm max}$, as it performs well in matching the full-physics radial profile and auto-correlation of the ELGs on scales of $r > 1 {\rm Mpc}/h$, or their dark-matter distribution, for which we randomly select particles in the halo.

\subsection{Constructing HOD catalogues}
\label{sec:mock.algo}
The most commonly-adopted formalism to building mock galaxy catalogues is the HOD.
In this section, we provide a prescription for implementing
environment assembly bias effects into mock catalogues utilizing
the HOD method and
test its efficacy by comparing the derived catalogues with the
``true'' TNG300 galaxies. The purpose of this is to demonstrate
that indeed incorporating halo environmental properties can help
recovered the observed clustering in hydro simulations. The
approach that we take utilizes halo occupation information 
as a function of both halo mass and environment as measured
directly in the simulation and follows a similar procedure
as that outlined in \cite{2020arXiv200705545X} and \citet{2020arXiv200503672C}. The benefit
of this method is that it is parameter-free and remains agnostic
about the particular shape of the HOD, which is advantageous
for any analyses working with non-standard occupation distributions.
As illustrated in Section \ref{sec:res.hod}, ELGs and
star-forming galaxies do not follow the traditional steadily
increasing HOD shape and thus require specialized modeling.

The steps for obtaining such a mock catalogue are outlined below:
\begin{itemize}
    \item[1.] We split the haloes belonging to each mass bin
    (of size $\Delta \log(M_{\rm halo}) = 0.1 \ {\rm dex}$) into 5
    bins based on their environment parameter (ranging from
    the 20\% haloes in the bin belonging to the least dense
    environments to the 20\% belonging to the most dense 
    environments), where we define environment as the Gaussian-smoothed matter density over a scale of $R_{\rm smooth} = 1.4 \ {\rm Mpc}/h$.
    \item[2.] Within each environment bin, we measure the average
    number of centrals and the average number of satellites
    and store the information in the form of a two-dimensional
    array per mass bin per environment bin.
    \item[3.] To construct the mock catalogue, for each halo in the
    dark-matter-only catalogue (TNG300-1-Dark), we ascribe it
    the appropriate number of central and satellite galaxies,
    drawing from a binomial and a Poisson distribution, respectively,
    with probability and mean determined by the
    particular mass bin and environment bin the halo belongs to.
    \item[4.] We populate the haloes on small scales by either a) assigning
    the new galaxy positions to the locations of the top subhaloes
    rank ordered by their $V_{\rm max}$ with a scatter of 
    $\log(V_{\rm max}) = 0.1$. This choice is informed
    by our finding that the number density profile of ELGs is
    traced well by that of the subhaloes with the highest values of $V_{\rm max}$ (see Fig.~\ref{fig:prof});
    b) ``painting'' galaxies on top of randomly selected particles in the halo. This approach is motivated by the left panels of Fig.~\ref{fig:prof}, where we found very good agreement between the ELG radial distribution and that of the dark matter particles in the halo, and is particularly useful in cases where we do not have subhalo catalogue information (which is becoming increasingly expensive to store with current simulation volumes).
\end{itemize}
We do not require that every halo that has galaxies hosts a central
galaxy since the ELG selection criteria disqualify quiescent large
centrals, which are often found in the centres of large haloes (see
Fig.~\ref{fig:hod}).

\begin{figure}
\includegraphics[width=1.\columnwidth]{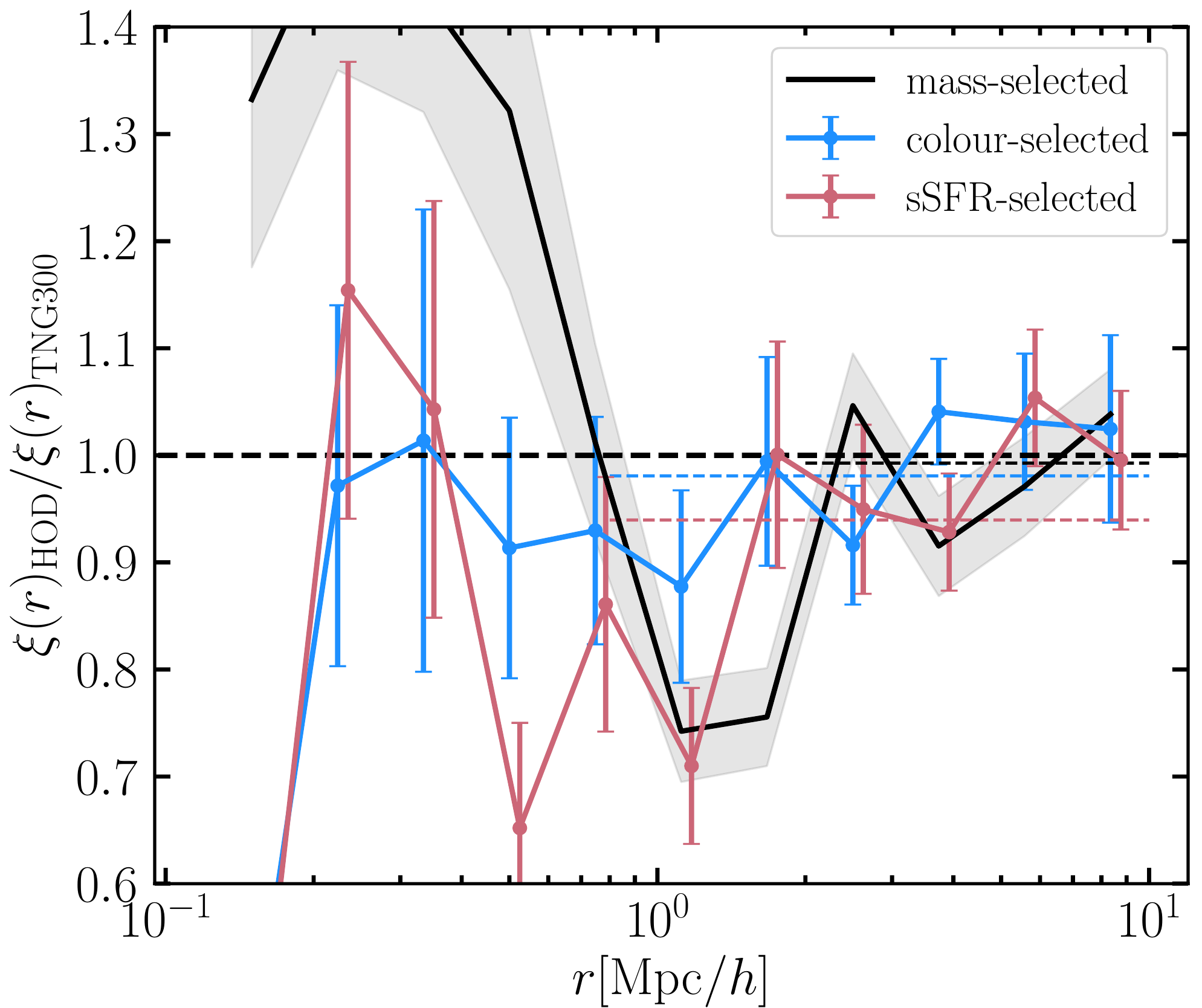} \\
\includegraphics[width=1.\columnwidth]{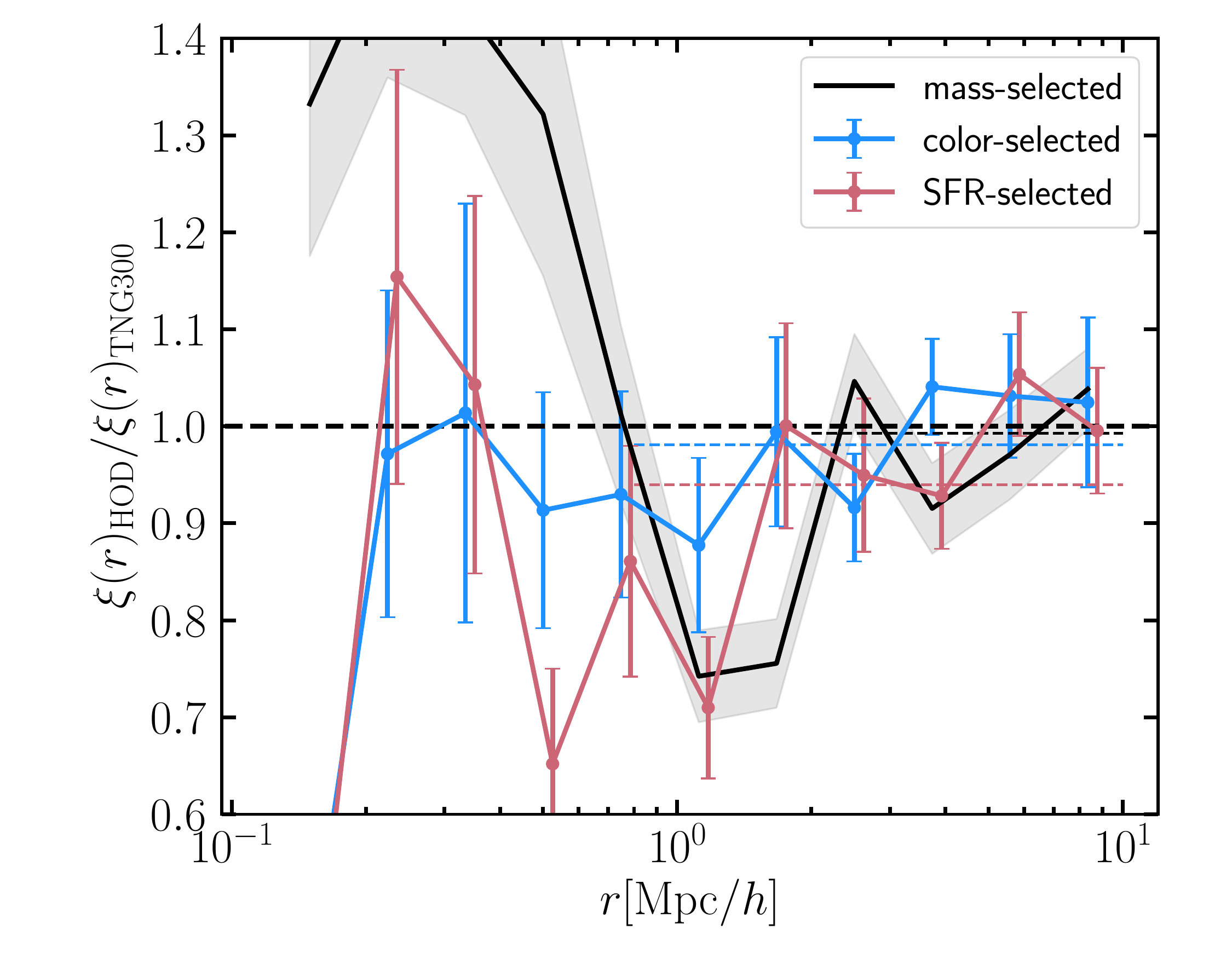}
\centering  
\caption{Clustering ratio between the two-dimensional HOD approach
augmented with environment information
described in Section \ref{sec:mock.algo} and the ``true'' galaxies for 
the following three samples: one selected based on a stellar-mass
cut (shown in \textit{grey solid}), one selected using the proposed
DESI cuts (shown in \textit{blue solid}), and one selected using the 
sSFR cuts (shown in \textit{red solid}). The \textit{top panel} shows a mock catalogue ratio obtained by using a rank-ordered list of the subhaloes based on $V_{\rm max}$ to assign the new galaxy locations, whereas in the \textit{bottom panel}, galaxies are ``painted'' on top of randomly selected particles in the halo (also see the left panels of Fig.~\ref{fig:prof}). We show that both methods produce congruent results, which is useful for creating mock catalogues when subhalo data are not available. In \textit{dotted lines},
we show the large-scale values for the three cases
averaged between $r = 0.8 \ {\rm Mpc}/h$ ($r = 2 \ {\rm Mpc}/h$ for the mass-selected sample) and 10 Mpc$/h$.
Compared with Fig.~\ref{fig:shuff}, we see that the discrepancies
with the ``true'' distributions are now smaller, which demonstrates
the efficacy of environment as an assembly bias parameter. On average
they deviate to within $\sim$1\% for the colour-selected ELGs and the
mass-selected samples, which is a much more palatable 
systematic effect to marginalize over in future mock challenges.}
\label{fig:mock}
\end{figure}

In Fig.~\ref{fig:mock}, we show the ratio of the clustering
between our mock catalogues and the ``true'' TNG300 galaxies at $z = 0.8$
for three scenarios. The \textit{top panel} shows the mock catalogue ratios obtained by using a rank-ordered list of the subhaloes based on $V_{\rm max}$ to assign the new galaxy locations, whereas in the \textit{bottom panel}, galaxies are ``painted'' on top of randomly selected particles in the halo (also see the left panels of Fig.~\ref{fig:prof}). We show that both methods produce congruent results, which is useful for creating mock catalogues when subhalo data are not available or not trusted (e.g., in the proximity of large clusters). In each of these scenarios, we use the 
two-dimensional HOD array derived from the corresponding
``true'' galaxy population to obtain the mock catalogues. 
For the first case, we compare the
mock catalogue with the TNG300 galaxies selected by applying the colour
selection criteria from the DESI survey. We observe that the
small discrepancy between the ``true'' and shuffled clustering
noted in Fig.~\ref{fig:mock} has been reconciled after
supplying the HOD model with knowledge about the halo environment.
Next, we compare the sample selected by
applying a sSFR cut (as described in
Section \ref{sec:meth.sfr} {and Table~\ref{tab:sfr}})
with the mock catalogue of that sSFR-selected sample, finding that the 
agreement between the two is again reasonable and
within the expected error margins of near-future galaxy surveys.
For the third case, we select galaxies
by applying a stellar mass cut chosen so as to keep the number
of galaxies equal to that in the other two cases (see Table \ref{tab:num}). 
We demonstrate that the mock catalogue for that sample recovers
successfully the clustering of the ``true'' galaxies on large scales,
significantly reducing the 10\% differences found when employing the mass-only
HOD formalism seen in Fig.~\ref{fig:shuff}.

\section{Discussion \& Conclusions}
\label{sec:disc}
Current and future cosmological surveys are actively targeting
luminous star-forming ELGs. In this paper, we
have studied how they populate the dark matter haloes formed 
inside the cosmic web structure using the state-of-the-art
hydrodynamical simulation IllustrisTNG. We model the
star-forming ELGs by imposing cuts in apparent magnitude
and colour-colour space in
an attempt to mimic the galaxy surveys eBOSS and DESI (see Table 
\ref{tab:obs}). An additional sample is obtained by applying
both a magnitude cut as well as a specific star-formation rate (sSFR)
threshold, which we have similarly designed to isolate
blue star-forming galaxies such as those targeted by the surveys
under consideration (see Fig.~\ref{fig:selection}). 
We have studied the spatial distribution of the 
DESI-selected ELG sample at $z = 0.8$ and compared it with
a stellar-mass-selected sample (see Fig.~\ref{fig:2d_distn}).
We have demonstrated that ELGs are likely to 
populate filamentary and sheet regions, defined via the traditional tidal
environment classification scheme (i.e. using the
eigenvalues of the tidal field to split the cosmic web into
peak, filaments, sheets, and voids). On the other hand, 
the galaxies in the mass-selected sample tend to reside in
the highest density regions (i.e. peaks, filaments).

Furthermore, we have shown that both the sSFR- 
and colour-selected samples behave very similarly when compared with 
each other in terms of their HODs and cumulative occupation distributions 
(see Fig.~\ref{fig:hod} and Fig.~\ref{fig:hod_manyz}). As found by
previous studies, we report that the occupation fraction
of central ELGs does not reach 100\% for any mass scale, but rather 
peaks around 10\% at low masses ($\log(M_{\rm halo}) \approx 12$).
Central ELGs tend not to
reside in more massive haloes, while the satellite contribution
follows the typical power law curve seen in traditional stellar-mass
selected samples.

We also compare the auto-correlation functions for the colour-
and sSFR-selected samples and find that their large-scale clustering
is in good agreement, with the largest differences occurring for 
the eBOSS redshift sample at $z = 0.8$ and the DESI redshift sample
at $z = 1.4$, both of which have a lower number of objects
(see Fig.~\ref{fig:corr} and Table \ref{tab:num}). The auto-correlation function on small
scales exhibits more notable discrepancy particularly for the 
$z = 0.8$ and $z = 1.1$ DESI samples. We then study their
relative clustering to a sample with shuffled halo occupations,
which mimics a mass-only HOD population model and find that
it affects ELG-like samples at about 4\% (see Fig.~\ref{fig:shuff}).
On the other hand, for mass-selected samples, the deviation is
around $\sim$10\% in agreement with previous works \citep[e.g.][]{2020MNRAS.493.5506H}.

We also study the large-scale bias and cross-correlation coefficient
of model ELGs and find it to be close to $\tilde b(r) \approx 1.4$ 
and roughly constant at $z = 0.8$ (see Fig.~\ref{fig:bias_corr}), 
while the cross-correlation coefficient $\tilde r(r)$ approaches 
unity around $r \sim~10 \ {\rm Mpc}/h$. This suggests 
that on these scales, baryon physics does not affect the
galaxy distribution, and the dominant source that governs
the galaxy distribution is gravity. We also find that the ELG-like
samples exhibit a much weaker bias compared with the mass-selected 
one. This result can be attributed to the fact that the mass-selected 
galaxies preferably populated the higher density regions, which are 
also more strongly biased (see Section \ref{sec:res.bias} for a 
discussion).

Many recent analyses have found that a halo parameter that
accounts for the majority of galaxy assembly bias effects
is halo environment. It has been shown that including 
it as an essential ingredient in mock recipes for mass-selected samples
is of utmost importance in order to model
galaxy clustering to a sufficiently high level of accuracy.
In this work, we define environment by applying a Gaussian smoothing
kernel to the dark matter field with a smoothing scale of 
$R_{\rm smooth} = 1.4 \ {\rm Mpc}/h$. By splitting the haloes
in each mass bin into those occupying low and high density 
regions, we have demonstrated (see Fig.~\ref{fig:hod_env})
that the occupation function is highly dependent on environment
both for the mass-selected sample as well as for our DESI
ELG-like sample, clearly showing that high-environment haloes
are more likely to be hosting ELGs particularly in the low halo-mass
regime. We also study the dependence of galaxy clustering
on environment, which is presented in Fig.~\ref{fig:corr_env},
and find that indeed galaxies living in high-environment haloes
are more clustered than their low-environment counterparts at 
fixed halo mass.

Implementing secondary properties into our HOD prescription
affects the two-halo term, as it changes the number of galaxies
within a halo at fixed halo mass and thus the pair counts, but 
it cannot inform us of sub-megaparsec processes. To construct
more accurate population models devoid of substantial inherent 
biases, it is also important to probe the galaxy-halo relationship
in the one-halo regime. In Fig.~\ref{fig:prof}, we explore several
different subhalo properties ($V_{\rm max}$, $V_{\rm peak}$, $V_{\rm disp}$, $M_{\rm SUBFIND}$, $M_{\rm max}$, and 
$M_{\rm twice}$) to use as proxies for determining
the placement of ELG satellites in both the full-physics and 
dark-matter-only runs of the TNG300 simulation box. In addition, we also display the density profile of haloes using dark matter particles and find that this provides a reasonably good fit to the ELG radial distribution. For the 
full-physics run, we find that all parameters perform reasonably
well with $V_{\rm max}$, $V_{\rm disp}$, and 
$M_{\rm twice}$ exhibiting the smallest amounts of discrepancy
for the clustering study (\textit{top right} panel). On the other
hand, in the dark-matter only simulation the satellite distribution
displays a lot more sizeable differences particularly on small
scales, where we see a notable flattening of the curves, which are attributable 
to the difference in the dynamics governing the two runs. Nevertheless,
we find that after adding a scatter of $\log(V) = 0.1$ to the velocity-based
parameters when performing the rank-ordering procedure,
we can recover the clustering better. The parameters providing the
best match are the velocity-based ones ($V_{\rm peak}$, $V_{\rm disp}$, and $V_{\rm max}$). In our recipe for constructing mock catalogues, we adopt $V_{\rm max}$ as well as a random particle selection method, which is useful in cases where we do not have reliable subhalo information.

Finally, we have applied a two-dimensional approach 
\citep[similar to][]{2020arXiv200705545X,2020arXiv200503672C} 
for incorporating halo environment
effects into HOD recipes and have shown that it manages to
successfully recover the hydro simulation clustering to within 1\%
(see Fig.~\ref{fig:mock}). This
is highly beneficial for improving the modeling of the galaxy-halo
relationship, which plays a central role in creating 
efficient mock catalogues. At their core, these mock catalogues tend 
to employ simple empirical population models and improving our models
could potentially greatly enhance our understanding
of the interaction between baryons and dark matter as well as
the systematic biases baked into them. Such an endeavor could 
thus bridge important gaps in light of on-going and future galaxy surveys.
Caveats of our analysis are the modest volume and resolution 
of the simulation we have employed as well as the dependence of our analysis on the particular subgrid model employed by the IllustrisTNG simulation. In particular, 
the latter could affect the SFR variability, leading to fewer star-busty 
systems than expected in reality. Regardless of these shortcomings, 
the main reason we study ELGs in the TNG hydrodynamical simulation is to
gesture towards possible directions for improving the
galaxy-halo connection model for ELGs rather than come up with a universal
prescription for ``painting'' galaxies to haloes.

In the near future, multiple surveys will be mapping the distribution of
emission-line galaxies, so understanding their clustering properties with
a high degree of fidelity will become crucial to advancing precision cosmology.
One of the most trusted methods for studying galaxy populations is through
hydrodynamical galaxy formation simulations. Therefore, the development of even 
larger boxes will be extremely beneficial for expanding our knowledge
of the relationship between galaxies and their dark 
matter haloes. Once such data sets become available,
we plan to test and validate the results obtained with TNG300
as well as improve the empirical population models used
for creating mock catalogues. Thanks to the substantially 
larger number of galaxies contained in these larger volume runs,
such simulations will enable us to capture the large-scale
behavior even better by possibly introducing a multidimensional
approach to the HOD model.

\section*{Acknowledgements}
We would like to thank Ben Johnson for providing us with useful insights regarding the stellar population synthesis code \textsc{python-fsps} as well as Lars Hernquist, Tanveer Karim and Sihan Yuan for their valuable input. S.T. is supported by the Smithsonian Astrophysical Observatory through the CfA Fellowship. S.B. is supported by Harvard University through the ITC Fellowship. D.J.E. is supported by U.S. Department of Energy grant DE-SC0013718 and as a Simons Foundation Investigator.

\section*{Data Availability}
The IllustrisTNG data is publicly available at
\url{www.tng-project.org}, while the scripts used in this
project are readily available upon request.




\bibliographystyle{mnras}
\bibliography{refs} 





\bsp	
\label{lastpage}
\end{document}